\documentclass[preprint,12pt]{elsarticle}

\usepackage{amsmath}
\usepackage{amssymb}
\usepackage{booktabs}
\usepackage{float}
\usepackage{graphicx}
\usepackage{hyperref}
\usepackage{subfigure}
\usepackage{tikz}
\usepackage{xcolor}

\journal{Computer Physics Communications}

\newcommand{\eps}{\epsilon}
\newcommand{\dd}{\mathrm{d}}
\newcommand{\vecJ}{\mathbf{J}}

\begin{document}

\begin{frontmatter}

\title{\texorpdfstring{\textsc{Chess}}{Chess}: CHEbyshev pSeudo-Spectral transport for Feynman integral differential equations}
\tnotetext[tnote1]{Report number: USTC-ICTS/PCFT-26-42} 

\author[addr1]{Yuanche Liu\texorpdfstring{\corref{cor1}\fnref{fn1}}{}}
\ead{liuyuanche@mail.ustc.edu.cn}

\author[addr1,addr2,addr3]{Yang Zhang\texorpdfstring{\corref{cor1}\fnref{fn2}}{}}
\ead{yzhphy@ustc.edu.cn}

\cortext[cor1]{Corresponding authors}
\fntext[fn1]{ORCID: \href{https://orcid.org/0009-0008-4604-1306}{0009-0008-4604-1306};
INSPIRE-HEP: \href{https://inspirehep.net/authors/2856186}{2856186}.}
\fntext[fn2]{ORCID: \href{https://orcid.org/0000-0001-9151-8486}{0000-0001-9151-8486};
INSPIRE-HEP: \href{https://inspirehep.net/authors/1062340}{1062340}.}

\address[addr1]{Interdisciplinary Center for Theoretical Study,
University of Science and Technology of China, Hefei, Anhui 230026, China}
\address[addr2]{Peng Huanwu Center for Fundamental Theory,
Hefei, Anhui 230026, China}
\address[addr3]{Center for High Energy Physics, Peking University,
Beijing 100871, People's Republic of China}

\begin{abstract}
We present \textsc{Chess} (CHEbyshev pSeudo Spectrum), a Wolfram Language
package for high-precision one-dimensional transport of \(\eps\)-factorized
differential equations for Feynman master integrals.  The solver works with the
matrix obtained by pulling a differential one-form to a chosen path.  This
matrix may be supplied directly, or assembled from constant matrices and
precomputed scalar pullbacks of the one-forms.  The program combines
Chebyshev--Lobatto spectral collocation, sparse matrix assembly, sequential
propagation in the \(\eps\)-expansion, and residue-based regularization of
spurious regular singular endpoints.  Benchmarks for large multi-scale integral
families show rapid node convergence and agreement with independent reference
data where such data are available.  In the fixed local-series comparison used
here, the Chebyshev transports also give shorter wall times; the reported
process-tree memory usage is comparable for the smaller parallel runs and lower
for the largest benchmark system in that comparison.

\medskip
\noindent\textbf{PROGRAM SUMMARY}

\begin{small}
\noindent
{\em Program Title:} \textsc{Chess} \\
{\em CPC Library link to program files:} to be added by Technical Editor \\
{\em Developer's repository link:} \url{https://github.com/Alice-Shimada/CHESS} \\
{\em Licensing provisions:} MIT license (MIT) \\
{\em Programming language:} Wolfram Language / Mathematica \\
{\em Supplementary material:} package file, examples, example data, sample outputs, and README \\
{\em Nature of problem:}\\
High-precision evaluation of Feynman master integrals by differential
equations requires transporting boundary values through large
\(\eps\)-factorized systems along paths in multi-scale kinematic space.  In
many applications the pulled-back coefficient matrix is a sparse combination
of constant matrices and scalar one-form pullbacks.  In other cases, including
systems with algebraic or elliptic ingredients, the same one-dimensional
transport problem can be posed by supplying the pulled-back matrix directly.
Local propagation methods can
require many expansion segments or small steps, especially when high precision
and endpoint behavior must be controlled. \\
{\em Solution method:}\\
\textsc{Chess} solves the pulled-back system by Chebyshev--Lobatto spectral
collocation.  The \(\eps\)-expanded hierarchy is solved sequentially, using
one factored scalar node matrix with many right-hand sides rather than a
monolithic node-by-integral system.  In the prepared-data mode, the package
caches the sparse constant-matrix decomposition of the differential equation
and evaluates the scalar one-form pullbacks only at collocation nodes.  In the
direct-matrix mode, the same propagator acts on a user-supplied numerical
evaluator for \(B(t)\).  Spurious regular singular endpoints are treated by
residue/finite-part extraction and row replacement or a lifted left-endpoint
unknown. \\
{\em Additional comments including restrictions and unusual features:}\\
The prepared-data interface accepts systems decomposed into constant matrices
times scalar differential one-forms, supplied after path pullback as
one-dimensional functions.  Logarithmic systems are a common special case, but
the spectral propagator only requires a numerical evaluator for the
one-dimensional \(\eps\)-factorized matrix \(B(t)\).  It can therefore also
be used for non-logarithmic examples once such an evaluator is supplied.
Endpoint regularization currently relies on residue and finite-part data for
the pulled-back matrix and is implemented most directly when the prepared
scalar pullbacks are available.  The program assumes that the basis, boundary
constants, path, and analytic branch have already been chosen by the user.
\end{small}
\end{abstract}


\end{frontmatter}

\section{Introduction}
\label{sec:introduction}

Multi-loop Feynman integrals are among the computational bottlenecks in
precision collider physics.  As scattering-amplitude calculations move to
higher loop orders, higher multiplicities, and increasingly differential
phenomenological applications, the relevant integral families involve many
kinematic scales, large sets of master integrals, and non-trivial singularity
structures.  Recent examples include three-loop five-point and two-loop
six-point massless systems, where the number of master integrals and the size
of the associated singular data make both numerical stability and computational
resource use important practical issues
\cite{Henn:2024ngj,Liu:2024ont,Chicherin:2025ThreeLoopFivePointComplete,Liu:2026TwoLoopSixPoint}.
From the point of view of computational physics, the same task has two faces:
one has to understand the analytic structure of the functions, and one has to
solve large structured linear systems at high precision.

Several complementary strategies are available for numerical Feynman-integral
evaluation.  Sector-decomposition and parametric-integration programs provide
general numerical access to dimensionally regularized integrals and continue
to evolve toward higher performance and broader physical-region coverage, as
illustrated for example by recent versions of \textsc{Fiesta},
\textsc{pySecDec}, and \textsc{feyntrop}
\cite{Smirnov:2021FIESTA5,Heinrich:2023PySecDec,Borinsky:2023Feyntrop}.
Another important direction is auxiliary-mass flow, implemented in
\textsc{AMFlow}, where the integrals are computed by solving differential
systems in an auxiliary mass parameter \cite{Liu:2022chg}.  Differential-equation
transport in kinematic variables is pursued by a number of public tools and
algorithms based on local or piecewise series solutions, including
\textsc{DiffExp}, \textsc{SeaSyde}, and \textsc{LINE}
\cite{Hidding:2020ytt,Armadillo:2022SeaSyde,Prisco:2025LINE}; see also the
recent review of differential-equation and series-expansion methods in
Ref.~\cite{Armadillo:2025Review}.  Related recent work includes direct
numerical integration of differential equations and the evaluation of iterated
integrals through differential-equation systems
\cite{PetitRosas:2026lgw,Baur:2026zlw}.  The package presented here belongs to this
general differential-equation ecosystem, but it targets a specific numerical
subproblem: high-precision one-dimensional transport of \(\eps\)-factorized
systems along a prescribed path.

The differential-equation method starts from a basis of master integrals and
expresses derivatives with respect to kinematic variables back in the same
basis by integration-by-parts identities and differential-equation techniques
\cite{Chetyrkin:1981qh,Laporta:2000dsw,Kotikov:1990kg,Remiddi:1997ny,Gehrmann:1999as}.
In a generic basis the system has the form
\begin{equation}
  \dd\mathbf{I}(\mathbf{x},\eps)
    =
  \mathcal{A}(\mathbf{x},\eps)\,
  \mathbf{I}(\mathbf{x},\eps),
  \label{eq:intro-generic-de}
\end{equation}
where \(\mathcal{A}\) is a matrix of differential one-forms.  In favorable
cases, a change of basis \(\vecJ=T(\mathbf{x},\eps)\mathbf{I}\)
brings the system to an \(\eps\)-factorized form
\cite{Henn:2013pwa,Lee:2014ioa,Meyer:2016slj,Prausa:2017ltv,Gituliar:2017vzm,Lee:2020zfb,e-collaboration:2025frv,Bree:2025tug}
\begin{equation}
  \dd\vecJ(\mathbf{x},\eps)
  =
  \eps\,\Omega(\mathbf{x})\,
  \vecJ(\mathbf{x},\eps).
  \label{eq:intro-epsilon-form}
\end{equation}
For many useful bases the matrix one-form can be written as a finite
constant-matrix decomposition
\begin{equation}
  \Omega(\mathbf{x})
  =
  \sum_{\alpha} A_{\alpha}\,\omega_{\alpha}(\mathbf{x}),
  \label{eq:intro-one-form-decomposition}
\end{equation}
where the matrices \(A_{\alpha}\) are constant and the \(\omega_{\alpha}\) are
scalar differential one-forms \cite{Henn:2013pwa,Lee:2014ioa}.  The usual
logarithmic representation is one special case of this notation.  Elliptic,
modular, or more general geometric sectors may involve one-forms whose
coefficients are algebraic functions or special functions
\cite{Frellesvig:2021hkr,Frellesvig:2023iwr,Gorges:2023zgv,Duhr:2024uid,Maggio:2025jel,Duhr:2025xyy}.
The spectral
transport step used below only needs the one-dimensional matrix obtained after
a path has been chosen.

Once boundary data are known at a point
\(\mathbf{x}(a)\), one may choose a path \(\mathbf{x}=\mathbf{x}(t)\),
\(t\in[a,b]\), and pull Eq.~\eqref{eq:intro-epsilon-form} back to a
one-dimensional ordinary differential equation.  Writing
\(\vecJ(t,\eps)=\vecJ(\mathbf{x}(t),\eps)\), this gives
\begin{equation}
  \frac{\dd}{\dd t}\vecJ(t,\eps)
  =
  \eps\,B(t)\vecJ(t,\eps),
  \qquad
  B(t)=\iota_{\dot{\mathbf{x}}(t)}\Omega(\mathbf{x}(t)).
  \label{eq:intro-pulled-back-B}
\end{equation}
For the decomposed one-form representation in
Eq.~\eqref{eq:intro-one-form-decomposition}, this becomes
\begin{equation}
  B(t)=\sum_{\alpha} A_{\alpha}\,b_{\alpha}(t),
  \qquad
  b_{\alpha}(t)=\iota_{\dot{\mathbf{x}}(t)}\omega_{\alpha}(\mathbf{x}(t)).
  \label{eq:intro-pulled-back-decomposition}
\end{equation}
In the logarithmic special case, \(b_{\alpha}(t)\) is the derivative along the
path of the corresponding logarithm.
The numerical problem is then a one-dimensional transport problem, but it is
not a small scalar ODE problem: the unknown is an \(\eps\)-expanded vector of
master integrals, the coefficient matrix may involve many scalar one-form
pullbacks or direct special-function evaluations, and the calculation is often
performed at high working precision.

This paper introduces \textsc{Chess} (CHEbyshev pSeudo Spectrum), a Wolfram
Language package for this one-dimensional transport problem.  It is useful to
distinguish two layers of the program.  The prepared-data interface assembles
the pulled-back matrix from constant matrices and precomputed scalar pullbacks
\(b_{\alpha}(t)\).  The underlying Chebyshev propagator, however, only requires a
numerical evaluator for \(B(t)\) in an \(\eps\)-factorized
one-dimensional system.  This lower-level interface is independent of whether
the entries of \(B(t)\) come from rational, logarithmic, elliptic, modular, or
other special-function data, provided they can
be evaluated reliably on the chosen collocation nodes.  The examples below use
both routes: the large multi-loop families use prepared scalar pullbacks, while
the non-planar triangle benchmark uses a direct matrix evaluator containing
elliptic periods from the published differential equation.

The package does not perform integral reduction, construct a canonical or
\(\eps\)-factorized basis, determine boundary constants, or select the
analytic-continuation path.  These data must be supplied by the user.  Within
this restricted scope, \textsc{Chess} makes repeated high-precision transports
less expensive while keeping the path data and analytic assumptions explicit.

The numerical method is Chebyshev--Lobatto pseudo-spectral collocation on the
transport interval.  Spectral collocation methods are standard in numerical
analysis \cite{Trefethen:2000Spectral,Canuto:2006Spectral}; their usefulness in
the present setting comes from analyticity in the complexified path parameter.
If the pulled-back matrix \(B(t)\) is analytic in a complex neighborhood of the
interval, then the transported solution is analytic there.  Equivalently, after
mapping the interval to \([-1,1]\), the convergence rate is controlled by the
largest Bernstein ellipse that avoids the nearest complex singularity of the
pulled-back connection.  In that situation, the Chebyshev coefficients decrease
geometrically, and the discretization error has the expected form
\(\mathcal O(\varrho^{-m})\) up to stability constants, where \(m\) is the
number of collocation nodes and \(\varrho>1\) is the corresponding Bernstein
parameter.  The nearest singularities are those of the pulled-back
connection.  In a prepared decomposition they are read from the scalar
pullbacks \(b_{\alpha}(t)\); in a direct evaluator they are singularities of the
supplied matrix-valued functions.

The \(\eps\)-factorized form gives another implementation advantage.
Expanding the solution in a finite Laurent window yields a triangular hierarchy
in the expansion order,
\begin{equation}
  \frac{\dd}{\dd t}\vecJ^{(r_{\min})}(t)=0,
  \qquad
  \frac{\dd}{\dd t}\vecJ^{(r)}(t)=B(t)\vecJ^{(r-1)}(t)
  \quad (r>r_{\min}).
  \label{eq:intro-epsilon-hierarchy}
\end{equation}
Thus each retained coefficient is obtained from the previous one by applying
the same one-dimensional connection and a spectral integration operator.  The
program uses this structure together with sparse matrix data, prepared scalar
pullbacks when available, and one factored scalar node matrix acting on many
right-hand sides.

Endpoint behavior is also important in practical transports.  Even if the
chosen path is regular in the physical region of interest, the one-dimensional
pullback may contain regular singular behavior at the left or right endpoint of
the computational interval.  When such an endpoint singularity is spurious for
the finite branch being transported, treating the coefficient matrix as an
ordinary smooth function can spoil a collocation solve.  \textsc{Chess}
implements residue-based left and right endpoint regularization for this case.  The
procedure extracts the pole and finite parts of the pulled-back connection and
replaces the endpoint equations by the corresponding finite regularity
conditions.  This is a numerical regularization layer for prescribed paths; it
is not an automatic branch-selection or path-planning algorithm.

The implementation follows the conventions of a CPC program paper: the source
code, example input data, command-line examples, and validation material are
provided with the program archive.  The package is written in Wolfram Language
and can be used from scripts or notebooks.  During the solve, it works with
preprocessed sparse matrices and numerical matrix values at collocation nodes,
rather than repeatedly manipulating large symbolic expressions.  The examples
test multi-scale integral families, check node refinement, compare with
independent reference data where available, and include a fixed-workflow
comparison with a local-series transport program.  The direct-matrix
non-planar triangle example uses the same propagator outside the prepared-data
workflow.

The rest of the paper is organized in the following way.
Section~\ref{sec:method}
introduces the Chebyshev collocation formulation and the endpoint
regularization.  Section~\ref{sec:implementation} describes the program
interface, prepared-data workflow, direct-matrix mode, and sparse assembly
strategy.  Section~\ref{sec:usage-validation-comparison} presents usage
examples, validation tests, the \textsc{DiffExp} comparison, and the
non-planar triangle direct-matrix benchmark.  Section~\ref{sec:conclusions}
summarizes the current capabilities and discusses possible extensions.

\section{Method}
\label{sec:method}

\subsection{\texorpdfstring{\(\eps\)-factorized systems and path pullback}{epsilon-factorized systems and path pullback}}
\label{subsec:formulation}

The numerical core of \textsc{Chess} solves an \(\eps\)-factorized
one-dimensional system
\begin{equation}
  \frac{\dd}{\dd t}\vecJ(t,\eps)
    =
  \eps\,B(t)\vecJ(t,\eps),
  \qquad t\in[a,b],
  \label{eq:one-dimensional-system}
\end{equation}
for a vector of master integrals or master-integral coefficients.  The matrix
\(B(t)\) may be supplied directly as a numerical evaluator.  Typically this
one-dimensional system is obtained by choosing a path in a multi-scale
kinematic space and pulling back an \(\eps\)-factorized differential
equation
\begin{equation}
  \dd\vecJ(\mathbf{x},\eps)
    =
  \eps\,\Omega(\mathbf{x})\,\vecJ(\mathbf{x},\eps).
  \label{eq:epsilon-factorized-system}
\end{equation}
The one-form matrix \(\Omega\) need not be logarithmic for the spectral solve
itself.  A useful prepared-data case is a finite decomposition into constant
matrices and scalar differential one-forms,
\begin{equation}
  \Omega(\mathbf{x})
    =
  \sum_{\alpha} A_{\alpha}\,\omega_{\alpha}(\mathbf{x}).
  \label{eq:one-form-decomposition}
\end{equation}
Here the matrices \(A_{\alpha}\) are constant, typically sparse matrices, and
the \(\omega_{\alpha}\) are scalar one-forms
\cite{Henn:2013pwa,Lee:2014ioa,Gehrmann:1999as,e-collaboration:2025frv,Bree:2025tug}.
Logarithmic systems are
included as the familiar special case
\(\omega_{\alpha}=\dd\log W_{\alpha}\).  The decomposition separates
the matrix data from the scalar functions that depend on the path, which is
the format used by the optimized prepared-data interface.  The construction of
the \(\eps\)-factorized basis, the reduction to master integrals, and the
determination of boundary constants are assumed to have been done before the
numerical transport begins.

After choosing a path \(\mathbf{x}=\mathbf{x}(t)\), pulling
Eq.~\eqref{eq:epsilon-factorized-system} back to the path gives
Eq.~\eqref{eq:one-dimensional-system} with
\begin{equation}
  B(t)=\iota_{\dot{\mathbf{x}}(t)}\Omega(\mathbf{x}(t)).
  \label{eq:path-pullback-general}
\end{equation}
For the decomposed one-form representation in Eq.~\eqref{eq:one-form-decomposition},
this becomes
\begin{equation}
  B(t)
    =
  \sum_{\alpha} A_{\alpha}\,b_{\alpha}(t),
  \qquad
  b_{\alpha}(t)=
    \iota_{\dot{\mathbf{x}}(t)}
      \omega_{\alpha}(\mathbf{x}(t)).
  \label{eq:path-pullback}
\end{equation}
In the prepared-data implementation, the scalar pullbacks are prepared before
the collocation solve:
\begin{equation}
  b_{\alpha}(t)
    =
  \iota_{\dot{\mathbf{x}}(t)}\omega_{\alpha}(\mathbf{x}(t)).
  \label{eq:prepared-scalar-pullbacks}
\end{equation}
The numerical matrix \(B(t)\) is then assembled from the prepared values
\(b_{\alpha}(t)\) and the sparse matrices \(A_{\alpha}\).  For a logarithmic
one-form this scalar function is
\(\dd\log W_{\alpha}(\mathbf{x}(t))/\dd t\).  This separation
keeps repeated numerical transports from redoing symbolic work along the path.
In the direct-matrix mode, this assembly layer is bypassed and a
user-supplied evaluator for \(B(t)\) is passed directly to the same spectral
propagator.

The \(\eps\)-factorized form gives a triangular hierarchy in the retained
Laurent coefficients.  We keep a finite Laurent window
\begin{equation}
  \vecJ(t,\eps)
    =
  \sum_{r=r_{\min}}^{r_{\max}}\eps^r \vecJ^{(r)}(t).
  \label{eq:epsilon-expansion}
\end{equation}
The implementation stores these coefficients by a non-negative column index,
but the recursion is independent of the absolute \(\eps\)-power offset.
Substituting Eq.~\eqref{eq:epsilon-expansion} into
Eq.~\eqref{eq:one-dimensional-system} gives
\begin{equation}
  \frac{\dd}{\dd t}\vecJ^{(r_{\min})}(t)=0,
  \qquad
  \frac{\dd}{\dd t}\vecJ^{(r)}(t)
    =
  B(t)\vecJ^{(r-1)}(t),
  \qquad r>r_{\min}.
  \label{eq:epsilon-hierarchy}
\end{equation}
Thus each coefficient \(\vecJ^{(r)}\) is obtained from the previous
coefficient by solving an inhomogeneous first-order system.  Unless an example
explicitly chooses a different orientation, \textsc{Chess} uses the convention
that the boundary coefficients \(\vecJ^{(r)}(a)\) are supplied at the
left endpoint and the transport proceeds from \(a\) to \(b\).

The method has a narrower scope than a complete
Feynman-integral computation framework.  The program takes as input a path,
boundary coefficients, and either prepared one-dimensional scalar pullbacks or
a direct numerical matrix evaluator.  It does not construct a canonical basis,
perform integration-by-parts reduction, infer boundary constants, select
analytic branches, or solve a genuinely multi-dimensional transport problem.

\subsection{Chebyshev spectral collocation}
\label{subsec:chebyshev}

After the pullback, the numerical task is to solve the triangular hierarchy on
the finite interval \([a,b]\).  \textsc{Chess} uses Chebyshev--Lobatto
collocation, following the standard spectral-method viewpoint that a regular
function on an interval can be represented accurately by its values on a
clustered polynomial grid, with exponential convergence when the function is
analytic in a suitable complex neighborhood
\cite{Trefethen:2000Spectral,Canuto:2006Spectral}.
On the reference interval the Lobatto nodes are
\begin{equation}
  x_j=\cos\frac{\pi j}{m},\qquad j=0,\ldots,m.
  \label{eq:lobatto-reference-nodes}
\end{equation}
They are mapped affinely to the transport interval,
\begin{equation}
  \widehat t_j
    =
  \frac{a+b}{2}+\frac{b-a}{2}x_j.
  \label{eq:transport-node-affine-map}
\end{equation}
The implementation stores the same set of nodes in increasing transport
direction, so that the ordered grid satisfies
\begin{equation}
  t_0=a,\qquad t_m=b.
  \label{eq:ordered-transport-endpoints}
\end{equation}
In what follows \(D\) denotes the correspondingly reordered and scaled
Chebyshev differentiation matrix on \([a,b]\).

For a vector-valued coefficient \(\vecJ^{(r)}\), differentiation at a
collocation node is approximated by
\begin{equation}
  \vecJ^{(r)\prime}(t_j)
    \approx
  \sum_{\ell=0}^{m}D_{j\ell}\vecJ^{(r)}(t_\ell).
  \label{eq:chebyshev-differentiation}
\end{equation}
In the ordinary case, where the left endpoint is not regularized by the lifted
unknown described below, the first row is replaced by the supplied boundary
condition and the differential equation is imposed at the remaining nodes:
\begin{equation}
  \vecJ^{(r)}(a)=\vecJ^{(r)}_a,
  \qquad
  \sum_{\ell=0}^{m}D_{j\ell}\vecJ^{(r)}(t_\ell)
    =
  B(t_j)\vecJ^{(r-1)}(t_j),
  \qquad j=1,\ldots,m.
  \label{eq:ordinary-collocation}
\end{equation}
The right-hand side is evaluated by assembling
\begin{equation}
  B(t_j)=\sum_{\alpha}A_{\alpha}\,b_{\alpha}(t_j)
  \label{eq:prepared-node-matrix}
\end{equation}
from the prepared scalar pullbacks.  Since the differentiation matrix and the
boundary row act only on the node index, the same scalar collocation operator
is reused for every master-integral component at a fixed \(\eps\) order.
The matrix-valued physics enters through the right-hand side
\(B(t_j)\vecJ^{(r-1)}(t_j)\), not through a new node operator for each
component.  This is the basic algebraic reason that the Chebyshev solve can be
organized as many right-hand sides for a factored node matrix rather than as a
large monolithic system over all nodes and all integrals.

The spectral discretization is global.  Changing one collocation value changes
the interpolating polynomial throughout the interval.  When the pulled-back
matrix \(B(t)\) is analytic on and near \([a,b]\), and when endpoint regular
singular behavior has been removed as described below, increasing the number of
nodes is expected to give spectral or near-spectral convergence.  This
expectation is not unconditional.  A true singularity on the path, an
unresolved branch choice, or a nearby complex singularity can reduce the
convergence rate or prevent convergence at the desired accuracy.  The practical
accuracy checks used in the paper compare results obtained from at
least two collocation grids, compare with independent reference values where
available, and otherwise use cross-method comparisons as consistency checks.

\subsection{Analyticity, Bernstein ellipses, and convergence rate}
\label{subsec:bernstein-convergence}

We now spell out the standard complex-analytic estimate behind the word
``spectral'' in the preceding subsection.  This discussion is included because
it gives a useful diagnostic for Feynman-integral transports: the observed
convergence rate is controlled not by the ordinary spectral radius of the
matrix on the real path, but by the nearest singularity of the pulled-back
one-dimensional connection in the complexified path parameter.  The classical
analysis of Chebyshev interpolation and pseudospectral differentiation may be
found, for example, in Refs.~\cite{Boyd:2001Spectral,Trefethen:2000Spectral,Canuto:2006Spectral,Trefethen:2019ATAP,Tadmor:1986Exponential,Demanet:2010ChebInterp}.

First map the transport interval to the standard Chebyshev interval,
\begin{equation}
  x=x(t)=\frac{2t-a-b}{b-a},
  \qquad
  t=t(x)=\frac{a+b}{2}+\frac{b-a}{2}x .
  \label{eq:t-x-map}
\end{equation}
For the purpose of the error analysis, write a scalar component of the exact
solution as a function of the reference variable, say
\(f(x)=(\vecJ^{(r)}(t(x)))_i\).  Chebyshev polynomials are defined by
\begin{equation}
  T_n(x)=\cos(n\arccos x),\qquad x\in[-1,1].
  \label{eq:chebyshev-polynomial-definition}
\end{equation}
Introducing
\begin{equation}
  z=e^{\mathrm{i}\theta},
  \qquad x=\cos\theta,
  \label{eq:z-theta-map}
\end{equation}
one obtains the Joukowski map
\begin{equation}
  x=\frac12\left(z+z^{-1}\right),
  \qquad
  T_n\!\left(\frac12(z+z^{-1})\right)
    =\frac12\left(z^n+z^{-n}\right).
  \label{eq:joukowski-chebyshev}
\end{equation}
This identity is the source of the Bernstein ellipse.  The unit circle
\(|z|=1\) is mapped to the real interval \([-1,1]\).  More generally, the
circle \(|z|=\varrho\), \(\varrho>1\), is mapped to the ellipse
\begin{equation}
  x(\theta)
    =
  \frac12\left(\varrho e^{\mathrm{i}\theta}
      +\varrho^{-1}e^{-\mathrm{i}\theta}\right),
  \qquad 0\leq\theta<2\pi .
  \label{eq:bernstein-parametric}
\end{equation}
Writing real and imaginary parts gives
\begin{equation}
  \operatorname{Re}x
    =\frac12(\varrho+\varrho^{-1})\cos\theta,
  \qquad
  \operatorname{Im}x
    =\frac12(\varrho-\varrho^{-1})\sin\theta .
  \label{eq:bernstein-real-imaginary-parts}
\end{equation}
Thus the image is the ellipse with foci at \(\pm1\), semi-major axis
\begin{equation}
  a_\varrho=\frac12(\varrho+\varrho^{-1}),
  \label{eq:bernstein-semi-major-axis}
\end{equation}
and semi-minor axis
\begin{equation}
  b_\varrho=\frac12(\varrho-\varrho^{-1}).
  \label{eq:bernstein-semi-minor-axis}
\end{equation}
This ellipse is denoted by \(E_\varrho\).  The parameter \(\varrho\) is often
called the elliptical radius.  Larger \(\varrho\) means that the function is
analytic in a wider complex neighborhood of the interval.

The coefficient estimate follows from the same map.  Suppose that \(f\) is
analytic in and on \(E_\varrho\), and define
\begin{equation}
  F(z)=f\!\left(\frac12(z+z^{-1})\right).
  \label{eq:cheb-F-definition}
\end{equation}
Then \(F\) is analytic in the annulus
\begin{equation}
  \varrho^{-1}<|z|<\varrho .
  \label{eq:cheb-annulus}
\end{equation}
The Chebyshev expansion, written out without the common primed-sum shorthand,
\begin{equation}
  f(x)=\frac{a_0}{2}+\sum_{n=1}^{\infty} a_n T_n(x)
  \label{eq:cheb-expansion}
\end{equation}
corresponds to the even Laurent expansion of \(F(z)\).  If
\begin{equation}
  M_r=\max_{x\in E_r}|f(x)|,
  \qquad 1<r<\varrho,
  \label{eq:cheb-Mr-definition}
\end{equation}
then Cauchy's estimate on the circle \(|z|=r\) gives the geometric coefficient
bound
\begin{equation}
  |a_n|\leq 2M_r r^{-n},
  \qquad n\geq 1 .
  \label{eq:cheb-coeff-bound}
\end{equation}
Since \(|T_n(x)|\leq1\) for \(x\in[-1,1]\), the tail of the Chebyshev series
satisfies
\begin{equation}
  \left\|f-\left(\frac{a_0}{2}+\sum_{n=1}^{m}a_nT_n\right)\right\|_{\infty,[-1,1]}
    \leq
  \sum_{n=m+1}^{\infty}|a_n|
    \leq
  \frac{2M_r}{r-1}\,r^{-m} .
  \label{eq:cheb-tail-bound}
\end{equation}
The same geometric rate applies to Chebyshev interpolants at Lobatto or
extremal points, up to the standard Lebesgue-constant factor, which grows only
logarithmically with \(m\).  Equivalently, for every \(1<\eta<\varrho\), the
algebraic or logarithmic prefactor can be absorbed into the constant, giving
\begin{equation}
  \|f-I_m f\|_{\infty,[-1,1]}\leq C_\eta \eta^{-m},
  \qquad 1<\eta<\varrho,
  \label{eq:cheb-interp-bound}
\end{equation}
where \(I_m f\) denotes the degree-\(m\) Chebyshev--Lobatto interpolant.  This
is the elementary origin of the exponential convergence used by the numerical
method.

The nearest singularity determines the largest admissible ellipse.  If
\(x_s\notin[-1,1]\) is a complex singularity of \(f\), the Bernstein ellipse
passing through \(x_s\) is obtained by solving the Joukowski equation
\begin{equation}
  x_s=\frac12\left(z+z^{-1}\right).
  \label{eq:singularity-joukowski-equation}
\end{equation}
Equivalently,
\begin{equation}
  z^2-2x_s z+1=0,
  \qquad
  z=x_s\pm\sqrt{x_s^2-1} .
  \label{eq:singularity-joukowski-roots}
\end{equation}
The two roots are reciprocal.  Hence the elliptical radius of the ellipse
through \(x_s\) is
\begin{equation}
  \varrho(x_s)=
  \max\left\{
  \left|x_s+\sqrt{x_s^2-1}\right|,
  \left|x_s-\sqrt{x_s^2-1}\right|
  \right\}.
  \label{eq:bernstein-parameter}
\end{equation}
For a singularity \(t_s\) in the original path parameter, one first maps it to
\begin{equation}
  x_s=\frac{2t_s-a-b}{b-a}
  \label{eq:singularity-reference-coordinate}
\end{equation}
and then uses Eq.~\eqref{eq:bernstein-parameter}.  The practical convergence
parameter is therefore
\begin{equation}
  \varrho_* = \min_{t_s}\varrho(t_s),
  \label{eq:rho-star}
\end{equation}
where the minimum runs over the singularities of the pulled-back coefficient
matrix, or of the solution itself, that are relevant for the chosen analytic
branch.  A rough asymptotic estimate is then
\begin{equation}
  E_m \sim C\,\varrho_*^{-m},
  \qquad
  \Delta \hbox{digits per node}\simeq \log_{10}\varrho_* .
  \label{eq:rho-star-error-trend}
\end{equation}
When a singularity approaches the interval, \(\varrho_*\to1\), and the
exponential advantage correspondingly deteriorates.

We now translate this scalar estimate to the differential-equation solve.  In
the reference variable the system may be written as
\begin{equation}
  \frac{\dd}{\dd x}\vecJ^{(r)}(x)
    =
  \widetilde B(x)\vecJ^{(r-1)}(x),
  \qquad
  \widetilde B(x)=\frac{b-a}{2}B(t(x)).
  \label{eq:reference-variable-system}
\end{equation}
If \(\widetilde B(x)\) is analytic in \(E_\varrho\), then the solution
coefficients are analytic there as well, after possibly reducing
\(\varrho\) so that the chosen branch contains no singularity.  This follows,
for example, from the equivalent Volterra equation and Picard iteration: an
integral of an analytic matrix times an analytic vector remains analytic, and
the iteration converges uniformly on compact subsets of the analytic domain.
Applying Eq.~\eqref{eq:cheb-interp-bound} componentwise gives
\begin{equation}
  \|\vecJ^{(r)}-I_m\vecJ^{(r)}\|_{\infty}
    \leq C^{(r)}_\eta\eta^{-m}.
  \label{eq:vector-chebyshev-interpolation-bound}
\end{equation}
Moreover, the derivative of an analytic Chebyshev approximation obeys the same
geometric rate with a different constant, again after replacing \(\varrho\) by
any \(\eta<\varrho\).  Consequently the residual of the interpolated exact
solution,
\begin{equation}
  \mathcal R_m^{(r)}(x)
    =
  \frac{\dd}{\dd x}I_m\vecJ^{(r)}(x)
  -\widetilde B(x)I_m\vecJ^{(r-1)}(x),
  \label{eq:interpolant-residual}
\end{equation}
satisfies
\begin{equation}
  \|\mathcal R_m^{(r)}\|_{\infty}\leq C^{(r)}_{\eta,\mathcal R}\eta^{-m}.
  \label{eq:residual-bound}
\end{equation}
The collocation solution is the polynomial for which the corresponding nodal
residual vanishes, together with the boundary row.  The actual error
is controlled by the residual bound multiplied by the stability constant of
the discrete collocation inverse.  For the regular path segments considered in
this paper this stability factor is monitored numerically by node refinement;
when it remains bounded, or grows at most algebraically with \(m\), the
geometric decay in Eq.~\eqref{eq:residual-bound} gives
\begin{equation}
  \|\vecJ^{(r)}-\vecJ^{(r)}_m\|_{\infty}
    \leq C^{(r)}_\eta\eta^{-m},
  \qquad 1<\eta<\varrho .
  \label{eq:collocation-error-bound}
\end{equation}
This is the sense in which the Chebyshev propagation has exponential error
control: the cost of increasing the node number is polynomial in \(m\), while
the analytic approximation error decreases geometrically until it reaches the
working precision or conditioning limits.

\subsection{Spurious endpoint regular singular points}
\label{subsec:endpoint}

The pullback to a one-dimensional path can introduce regular singular behavior
at an endpoint even when the transport problem itself has a regular limiting
value there.  Such endpoint factors are inconvenient for a collocation method:
the coefficient matrix \(B(t)\) is not finite at the endpoint, although the
solution may be.  The endpoint regularization in \textsc{Chess} is designed
for this spurious regular singular case.  It is not an analytic-continuation
prescription for crossing physical singularities.  The need to impose
endpoint compatibility conditions in singular boundary-value and spectral
collocation problems is standard in numerical analysis
\cite{Keller:1976SingularProblems,Huang:2003SpectralSingular}; the particular
finite-source recursion below follows from the
\(\eps\)-triangular hierarchy in Eq.~\eqref{eq:epsilon-hierarchy}.

Let \(c\in\{a,b\}\) be an endpoint at which \(B(t)\) has a one-sided regular
singular expansion.  In the global \(t\)-coordinate convention we write
\begin{equation}
  B(t)=\frac{R_c}{t-c}+M_c+O(t-c),
  \label{eq:endpoint-expansion}
\end{equation}
where \(R_c\) is the residue matrix and \(M_c\) is the finite part.  The
boundary data at an active regularized endpoint are assumed to satisfy the
regularity condition
\begin{equation}
  R_c\vecJ^{(r)}(c)=0
  \label{eq:residue-regularity}
\end{equation}
for the relevant \(\eps\)-coefficients.  This condition removes the
non-integrable endpoint term in the inhomogeneous hierarchy.  At a left
regularized endpoint this condition is checked directly on the supplied
boundary data.  At a right regularized endpoint it is a consistency condition
on the endpoint values obtained at lower retained orders, because the finite
source for order \(r\) depends on \(\vecJ^{(r-1)}\).  Regularity of the
highest retained order would only be needed to form the finite source for the
next order.

For the triangular equation
\begin{equation}
  \vecJ^{(r)\prime}(t)=B(t)\vecJ^{(r-1)}(t),
  \label{eq:triangular-endpoint-equation}
\end{equation}
the finite endpoint source is obtained by expanding the right-hand side at
\(t=c\).  In the endpoint formulas \(r\) denotes the non-negative index after
shifting the lowest retained Laurent coefficient to \(r=0\); the source for
that lowest coefficient is zero.  If
\begin{equation}
  \vecJ^{(r-1)}(t)
    =
  \vecJ^{(r-1)}(c)+(t-c)S_{r-1}(c)+O((t-c)^2),
  \label{eq:endpoint-lower-order-expansion}
\end{equation}
then substituting Eq.~\eqref{eq:endpoint-expansion} into
Eq.~\eqref{eq:triangular-endpoint-equation} gives the finite value
\begin{equation}
  S_r(c)
    =
  M_c\vecJ^{(r-1)}(c)+R_cS_{r-1}(c),
  \qquad
  S_0(c)=0.
  \label{eq:endpoint-source-recursion}
\end{equation}
Equivalently,
\begin{equation}
  S_r(c)
    =
  \sum_{q=1}^{r}R_c^{q-1}M_c\vecJ^{(r-q)}(c).
  \label{eq:endpoint-source-sum}
\end{equation}
The implementation evaluates this expression sequentially: it first forms
\(M_c\vecJ^{(r-q)}(c)\), then applies powers of \(R_c\).  This order is
important in sparse examples because it preserves structural zeros that may be
obscured by precomputing composite products.

At a regularized right endpoint, the left boundary row remains unchanged.  The
collocation equation is imposed at the interior nodes, while the right endpoint
differential row is replaced by the finite source condition
\begin{equation}
  \sum_{\ell=0}^{m}D_{m\ell}\vecJ^{(r)}(t_\ell)
    =
  S_r(b).
  \label{eq:right-endpoint-row}
\end{equation}
This keeps the unknowns as the endpoint values
\(\vecJ^{(r)}(t_\ell)\), but avoids evaluating the singular expression
\(B(b)\vecJ^{(r-1)}(b)\).

The left endpoint requires a different treatment because the same node would
otherwise have to carry two incompatible roles: it must impose the supplied
boundary value and it must also replace a singular differential equation by its
finite limit.  \textsc{Chess} avoids this conflict by removing the known
boundary value before constructing the collocation system.  It writes
\begin{equation}
  \vecJ^{(r)}(t)
    =
  \vecJ^{(r)}_a+(t-a)\mathbf{Z}^{(r)}(t),
  \qquad
  \vecJ^{(r)}_a=\vecJ^{(r)}(a).
  \label{eq:left-lift}
\end{equation}
Thus \(\mathbf{Z}^{(r)}\) is the regular one-sided difference quotient
\((\vecJ^{(r)}(t)-\vecJ^{(r)}_a)/(t-a)\).  If the transported branch
is finite at \(a\), then \(\mathbf{Z}^{(r)}(a)\) is exactly the one-sided
derivative \(\vecJ^{(r)\prime}(a)\).  Differentiating
Eq.~\eqref{eq:left-lift} gives
\begin{equation}
  \vecJ^{(r)\prime}(t)
    =
  \mathbf{Z}^{(r)}(t)+(t-a)\mathbf{Z}^{(r)\prime}(t),
  \label{eq:left-lift-derivative}
\end{equation}
so the scalar node operator acting on the values of
\(\mathbf{Z}^{(r)}\) is
\begin{equation}
  L_{j\ell}=\delta_{j\ell}+(t_j-a)D_{j\ell}.
  \label{eq:left-lift-operator}
\end{equation}
Away from the left endpoint, the lifted collocation equation becomes
\begin{equation}
  \sum_{\ell=0}^{m}L_{j\ell}\mathbf{Z}^{(r)}(t_\ell)
    =
  B(t_j)\vecJ^{(r-1)}(t_j),
  \qquad j=1,\ldots,m,
  \label{eq:left-lift-collocation}
\end{equation}
with the last index omitted from this range when the right endpoint is also
regularized.  The left endpoint row is not evaluated with \(B(a)\).  It is the
finite derivative condition
\begin{equation}
  \mathbf{Z}^{(r)}(a)=S_r(a).
  \label{eq:left-endpoint-row}
\end{equation}
The source \(S_r(a)\) is the finite limit of
\(B(t)\vecJ^{(r-1)}(t)\) obtained from
Eq.~\eqref{eq:endpoint-source-recursion}.  Once
\(\mathbf{Z}^{(r)}\) has been found, the original coefficient is recovered by
Eq.~\eqref{eq:left-lift}.  This is still a Chebyshev collocation solve: only the
scalar node operator changes from \(D\) to \(L\), while the matrices
\(A_{\alpha}\), \(R_a\), and \(M_a\) enter through the ordinary right-hand side
and the finite endpoint source.

The sign convention for residues is fixed as follows.  At the left endpoint
one expands in \(t=a+s\).  At the right endpoint one may compute a one-sided
expansion in \(t=b-s\), but the residue is converted back to the global
\((t-c)^{-1}\) convention used in Eq.~\eqref{eq:endpoint-expansion}.  With
this convention, the same formula for \(S_r(c)\) applies to both endpoints.

\subsection{Accuracy expectations and failure modes}
\label{subsec:accuracy}

The accuracy of the method is assessed through convergence and comparison
rather than through a single discretization.  In the validation section, one
should compare a coarser and a finer Chebyshev grid at the same working
precision, compare with independent reference values whenever such values are
available, and use independent implementations as cross-checks when no
higher-precision reference is available.  Agreement between two node counts is
evidence that the global polynomial approximation has stabilized on the chosen
path.  Agreement with an independent implementation checks branch conventions,
boundary data, and implementation details, although it is not by itself a
mathematical certificate of absolute accuracy.

The working precision must be chosen above the requested output accuracy.
Several effects consume guard digits: the conditioning of the collocation
operator, numerical extraction of endpoint residues and finite parts,
cancellation in the \(\eps\)-order recursion, and multiplication by sparse
but large matrices along the path.  For this reason, a failed high-accuracy run
does not by itself identify a flaw in the method.  It may indicate that the
node count, working precision, endpoint data, or path choice must be adjusted.

The most important failure modes are geometric or analytic.  The path may
cross a true singular hypersurface of the original differential system.  An
endpoint pole may be physical rather than spurious, so that the assumed
regularity condition \(R_c\vecJ^{(r)}(c)=0\) is not satisfied.  A prepared
scalar pullback may use the wrong branch along the path.  The supplied
boundary constants may be inconsistent with the residue constraints.  Even
when the real interval is free of singularities, a nearby complex singularity
can require more nodes than expected, and insufficient working precision can
hide an otherwise convergent spectral approximation.

These limitations define the intended scope of \textsc{Chess}.  The package
does not automatically deform paths around singularities, construct boundary
data, choose analytic branches without user input, or turn a generic system
into an \(\eps\)-factorized one.  Within its intended setting, namely a
fixed one-dimensional \(\eps\)-factorized transport problem supplied either
through prepared scalar pullbacks or through a direct matrix evaluator, the
method works best when the path data are analytic on a neighborhood of the
interval and any endpoint singularities are spurious regular singular points
that satisfy the residue regularity conditions.

\section{Implementation and package interface}
\label{sec:implementation}

The release file is named \texttt{Chess.wl}.  It implements the
one-dimensional transport problem described in Sec.~\ref{sec:method}; it does
not contain case-specific loaders, data conversion code, boundary construction,
or integral reduction.  In a typical workflow, a notebook or preprocessing
script converts the chosen family to one of the input conventions used by the
package, and \textsc{Chess} then evaluates the one-dimensional matrix and
solves the Chebyshev collocation problem.

\subsection{Input data model}
\label{subsec:implementation-input}

The package assumes that the differential equation has already been restricted
to a path in kinematic space.  There are two input layers.  In the
prepared-data mode, the user supplies
\texttt{Atilde}, \texttt{dLettersLine}, and the boundary matrix \texttt{y0}.  In
the direct-matrix mode, the user supplies a numerical Wolfram Language function
which returns the pulled-back matrix \(B(t)\) directly, together with the same
boundary matrix and interval.

In the prepared-data mode, \texttt{Atilde} is a formal slot expression written
with the unified markers \texttt{Log[W[i]]} or \texttt{logW[i]}.  The names
come from the first applications that motivated the interface, but in
the solver they only label scalar one-form pullbacks.  In the notation of
Sec.~\ref{subsec:formulation}, \texttt{Atilde} contains the constant matrices
\(A_{\alpha}\), and the integer index \texttt{i} labels the slot
\(\alpha\).  The object \texttt{dLettersLine} is a list of the prepared
one-dimensional functions
\begin{equation}
  b_{\alpha}(t)=
  \iota_{\dot{\mathbf{x}}(t)}\omega_{\alpha}(\mathbf{x}(t)).
  \label{eq:implementation-scalar-pullback-input}
\end{equation}
If there are \(n_{\mathrm{MI}}\) master integrals and \(n_\eps\) retained
\(\eps\)-coefficients, the boundary matrix \texttt{y0} has dimension
\(n_{\mathrm{MI}}\times n_\eps\).  The transport interval is supplied as
the list \(\{x0,x1\}\).

The standard evaluator for the prepared-data matrix \(B(t)\) is
\texttt{nAt}.  Once \texttt{Atilde} and \texttt{dLettersLine} have been defined,
\texttt{nAt[t, "Precision" -> p]} substitutes the numerical value of \(t\)
into the prepared scalar pullbacks and returns the sparse matrix \(B(t)\) at the
requested precision.  This helper is intentionally thin: family-dependent
choices, such as how to read boundary constants or how to map a particular
data format to the slot markers, are kept outside the package core.

The direct-matrix mode bypasses \texttt{Atilde} and \texttt{dLettersLine}.  The
first argument of \texttt{SpectralPropagate} may be any function that returns a
numerical matrix \(B(t)\) at the requested precision.  This interface is used
for the non-planar triangle benchmark in Sec.~\ref{subsec:np-triangle-direct},
where the pulled-back matrix is evaluated directly from the published
differential-equation matrix rather than from a prepared slot decomposition.

\subsection{Prepared-data workflow}
\label{subsec:implementation-prepared}

Internally, \textsc{Chess} decomposes \texttt{Atilde} into sparse slot
matrices,
\begin{equation}
  \sum_{\alpha} A_{\alpha}\,q_{\alpha}.
  \label{eq:implementation-slot-decomposition}
\end{equation}
Here \(q_{\alpha}\) is only a formal marker for a scalar pullback slot.  Terms
independent of the markers do not contribute to the differential-matrix
assembly.  After \texttt{dLettersLine} is evaluated at a node, the package forms
\begin{equation}
  B(t)=\sum_{\alpha} A_{\alpha}\,b_{\alpha}(t) .
  \label{eq:implementation-prepared-assembly}
\end{equation}

The helper routines with prefix \texttt{CHESS} prepare and cache this sparse
slot representation.  The main names are
\begin{verbatim}
CHESSNormalizeAtildeInput
CHESSBuildAtildeLinearData
CHESSAtildeLinearData
CHESSAtildeLinearCombination
\end{verbatim}
They are mentioned here only to make the data flow transparent; ordinary users
usually interact with them through \texttt{nAt}.  The performance-critical
representation is based on plain lists and \texttt{SparseArray} objects rather
than Mathematica \texttt{Association} objects.

\subsection{Main solver}
\label{subsec:implementation-solver}

The public solver call has the form
\begin{verbatim}
result = SpectralPropagate[
  nAt,
  y0,
  {x0, x1},
  "Nodes" -> m,
  "Precision" -> prec,
  "WorkingPrecisionA" -> precA,
  "RegularizedEndpoints" -> spec,
  "ParallelEvaluation" -> Automatic,
  "ParallelKernels" -> kernels
];
\end{verbatim}
The first argument is a matrix evaluator.  In the prepared-data workflow this
is normally \texttt{nAt}; in the direct-matrix workflow it can be any user
function returning the numerical matrix \(B(t)\).  The second argument is the
boundary matrix, and the third argument is the interval in the path parameter.
The solver currently uses the sequential \(\eps\)-weight strategy described
in Sec.~\ref{subsec:chebyshev}: the coefficient \(\vecJ^{(r)}\) is solved
after \(\vecJ^{(r-1)}\) has been computed.

At each \(\eps\) order, the Chebyshev node matrix is scalar in the
master-integral index.  \textsc{Chess} factors the node-space
collocation matrix once and applies the resulting linear solver to many
right-hand sides, one for each master-integral component.  This avoids
materializing a large monolithic node-by-integral matrix.

\begin{table}[t]
\centering
\caption{Main options of \texttt{SpectralPropagate}.}
\label{tab:solver-options}
\scriptsize
\setlength{\tabcolsep}{4pt}
\renewcommand{\arraystretch}{1.12}
\begin{tabular}{@{}p{0.40\linewidth}@{\hspace{1em}}p{0.52\linewidth}@{}}
\toprule
Option & Meaning \\
\midrule
\texttt{"Nodes"} & Number \(m\) of Chebyshev subintervals, giving \(m+1\) Lobatto nodes. \\
\addlinespace[0.25em]
\texttt{"Precision"} & Precision used for collocation matrices and solution values. \\
\addlinespace[0.25em]
\texttt{"WorkingPrecisionA"} & Precision used when evaluating \(B(t)\). \\
\addlinespace[0.25em]
\texttt{"EndpointRegularizationPrecision"} & Precision for endpoint residue and finite-part extraction; \texttt{Automatic} uses \texttt{"WorkingPrecisionA"}. \\
\addlinespace[0.25em]
\texttt{"RegularizedEndpoints"} & Endpoint specification: \(\{\}\), \texttt{None}, \texttt{"Left"}, \texttt{"Right"}, \texttt{All}, or explicit endpoint coordinates. \\
\addlinespace[0.25em]
\texttt{"ParallelEvaluation"} & \texttt{Automatic}, \texttt{True}, or \texttt{False} for parallel preparation of independent node and endpoint data. \\
\addlinespace[0.25em]
\texttt{"ParallelKernels"} & Defaults to \(1\); an integer requests a fixed number of subkernels, while \texttt{All} or \texttt{Infinity} requests the local Mathematica limit. \\
\addlinespace[0.25em]
\texttt{"LinearSolverMethod"} & Passed to \texttt{LinearSolve} when not \texttt{Automatic}. \\
\addlinespace[0.25em]
\texttt{"SolveStrategy"} & Compatibility option; the implemented strategy is \texttt{"SequentialEpsilon"}. \\
\bottomrule
\end{tabular}
\end{table}

\subsection{Endpoint handling and diagnostics}
\label{subsec:implementation-endpoint}

Endpoint regularization uses the prepared scalar pullbacks when they are
available.
At a regularized endpoint \(c\), the package extracts the residue and finite
matrices \(\{R_c,M_c\}\) from the one-sided expansion of \(B(t)\).  For each
prepared scalar pullback, \textsc{Chess} first tries direct endpoint
substitution.  Only pullbacks that are singular at the endpoint require Laurent
coefficients, which are obtained with \texttt{SeriesCoefficient}.  The sparse pair
\(\{R_c,M_c\}\) is then assembled using the same slot matrices used by
\texttt{nAt}.  Direct-matrix examples can use the ordinary solver without this
prepared endpoint extraction; endpoint regularization in that mode
requires the user to supply equivalent residue and finite-part information or a
matrix evaluator from which these data can be extracted.

The option \texttt{"RegularizedEndpoints"} controls where this replacement is
active.  The value \texttt{"Left"} refers to the left interval endpoint
\texttt{x0}, \texttt{"Right"} refers to \texttt{x1}, and \texttt{All}
activates both.  As in Sec.~\ref{subsec:endpoint}, a right endpoint replaces
the endpoint derivative row by the finite source \(S_r(b)\), while a left
endpoint uses the lifted unknown \(\mathbf{Z}^{(r)}\).

Two diagnostics are intended for routine use.  The function
\begin{verbatim}
EndpointRegularityResidual[a, yAtEndpoint, {x0, x1}, prec]
\end{verbatim}
checks the size of \(R_a y(a)\) for supplied endpoint data.  After a solve,
\begin{verbatim}
CHESSEndpointInfo[result]
\end{verbatim}
reports endpoint-preparation information, including how many scalar pullbacks
were evaluated by direct substitution and how many required Laurent
coefficients.
\texttt{CHESSClearCaches[]} starts a fresh problem without reloading the
package.

\subsection{Parallel execution}
\label{subsec:implementation-parallel}

Parallelism is used only where the work is independent.  The two parallel
stages are endpoint scalar-pullback extraction and ordinary node evaluation of
\(B(t_j)\).  The \(\eps\)-layers themselves are not parallelized, because
the layer \(r\) right-hand side depends on the already computed layer
\(r-1\).

The default is conservative: \texttt{"ParallelKernels" -> 1}.  An integer
value requests a fixed number of local subkernels.  The values \texttt{All}
and \texttt{Infinity} request the largest local subkernel count exposed by the
Mathematica configuration, and the package caps the request at that local
limit.  Existing kernels are reused when possible; missing kernels are
launched only when parallel evaluation is active.  The package distributes the
prepared scalar data and cached matrix decomposition to subkernels before
parallel node evaluation.

This design does not assume linear speedup with the number of kernels.  Its
purpose is to remove independent setup bottlenecks without changing the
sequential mathematical dependency between \(\eps\)-orders.  Benchmark
tables in Sec.~\ref{sec:usage-validation-comparison} state the
kernel count explicitly.

\subsection{Result accessors}
\label{subsec:implementation-results}

\texttt{SpectralPropagate} returns a compact result object.  Users should read
it through accessors rather than direct list indexing:
\begin{itemize}
\item \texttt{CHESSNodes[result]}\\
  returns the ordered Chebyshev--Lobatto nodes.
\item \texttt{CHESSStateVectors[result]}\\
  returns the per-node solution data in
  the legacy flattened order used by existing scripts.
\item \texttt{CHESSFinalState[result]}\\
  returns the
  \(n_{\mathrm{MI}}\times n_\eps\) matrix at the right endpoint, which is
  the most common object for validation and downstream use.
\item \texttt{CHESSEndpointSubBlocks[result]}\\
  returns the endpoint operator data used by the solve.
\item \texttt{CHESSEndpointInfo[result]}\\
  returns endpoint diagnostics.
\item \texttt{CHESSResultPart[result, key]}\\
  provides a uniform keyed accessor for the same fields.
\end{itemize}

\subsection{Implementation choices}
\label{subsec:implementation-performance}

The implementation is tuned for large one-dimensional transports.  The
important choices are sparse positional storage for the slot matrices, cache
reuse for repeated node evaluations, sparse endpoint residue and finite
matrices, sequential \(\eps\)-recursion with one factored scalar node
matrix, and parallel preparation for independent node and endpoint tasks.  In
particular, performance-critical data structures are lists and sparse arrays
rather than keyed containers.  These choices are specific to the transport
problem targeted by \textsc{Chess}; concrete timings, memory measurements, and
node-convergence tests are given in Sec.~\ref{sec:usage-validation-comparison}.

\section{Program usage, validation, and benchmarks}
\label{sec:usage-validation-comparison}

This section collects three benchmark groups.  The first is a physical-region
two-loop six-point double-pentagon transport checked against independent
\textsc{AMFlow} data.  The second uses the four large planar three-loop
five-point families to test the endpoint-regularized prepared-data workflow;
within this group we also include a fixed-protocol comparison with
\textsc{DiffExp}.  The third uses a two-loop non-planar triangle family with
elliptic periods to exercise the direct-matrix-evaluator mode on a system whose
letters are not restricted to ordinary logarithmic forms.

All timings in this section were obtained on a workstation with a 13th Gen
Intel Core i9-13950HX processor (24 physical cores, 32 hardware threads) and
128 GB of RAM, running Ubuntu 22.04.5 LTS.  We used Wolfram Mathematica 14.3
and requested 24 local parallel kernels.  In all reported runs the requested
number of kernels was also the number actually used.  Memory is reported as
the peak resident set size of the whole process tree, measured by an external
sampler rather than Mathematica's internal \texttt{MaxMemoryUsed[]} counter.

\subsection{Physical-region two-loop six-point DP benchmark}
\label{subsec:dp-physical}

The first benchmark revisits the physical-region double-pentagon example of
Ref.~\cite{Liu:2026TwoLoopSixPoint}.  The auxiliary files contain the
one-dimensional scalar pullbacks, the high-weight DP differential-equation
matrix, and two \textsc{AMFlow} numerical points in the physical region.  The
transport starts from \(\mathbf{x}_0\) and ends at the nearby point \(\mathbf{x}_1\),
with
\begin{equation}
\begin{aligned}
\mathbf{x}_0={}&
\left(\frac{325}{16},-\frac{181}{16},\frac{65}{8},\frac{9}{4},
\frac{19}{8},-\frac{19}{16},7,-\frac{109}{16},\frac{115}{8}\right),\\
\mathbf{x}_1={}&
\left(\frac{41202613}{1969920},-\frac{433189109}{37696000},
\frac{5365589}{634880},\frac{403880939}{152668800},
\right.\\
&\left.\hspace{2.6em}
\frac{57623740097}{24427008000},-\frac{1115291}{1036800},
\frac{10965881791}{1526688000},\right.\\
&\left.\hspace{2.6em}
-\frac{2160073789}{301568000},
\frac{30049703}{1969920}\right).
\end{aligned}
\label{eq:dp-physical-points}
\end{equation}
Here the 9 kinematic variables are selected as: 
\begin{equation}
    \mathbf{x} = \left(s_{12},\,s_{23},\,s_{34},\,s_{45},\,s_{56},\,s_{61},\,s_{123},\,s_{234},\,s_{345}\right)
\end{equation}
where $s_{ij} \equiv (p_i+p_j)^2$ and $s_{ijk} = (p_i+p_j+p_k)^2$.
Both endpoints lie in the same physical scattering region and no endpoint
regularization is used.  We compare the complete transported
\(\eps\)-coefficient matrix, with coefficients \(\eps^{-4}\) through
\(\eps^2\), against the independent \textsc{AMFlow} value at
\(\mathbf{x}_1\).

\begin{figure}[H]
\centering
\begin{tikzpicture}[x=1cm,y=1cm, line cap=round, line join=round,
  prop/.style={thick}, ext/.style={thick}, v/.style={circle,fill,inner sep=1.1pt},
  every node/.style={font=\scriptsize}]
  \coordinate (T) at (0,0.82);
  \coordinate (B) at (0,-0.82);
  \coordinate (L1) at (-1.35,0.92);
  \coordinate (L2) at (-1.95,0);
  \coordinate (L3) at (-1.35,-0.92);
  \coordinate (R1) at (1.35,0.92);
  \coordinate (R2) at (1.95,0);
  \coordinate (R3) at (1.35,-0.92);
  \draw[prop] (L1)--(T)--(B)--(L3)--(L2)--cycle;
  \draw[prop] (R1)--(T)--(B)--(R3)--(R2)--cycle;
  \draw[prop] (T)--(B);
  \draw[ext] (L1)--++(-0.75,0.55) node[left] {$p_1$};
  \draw[ext] (L2)--++(-0.9,0) node[left] {$p_2$};
  \draw[ext] (L3)--++(-0.75,-0.55) node[left] {$p_3$};
  \draw[ext] (R1)--++(0.75,0.55) node[right] {$p_4$};
  \draw[ext] (R2)--++(0.9,0) node[right] {$p_5$};
  \draw[ext] (R3)--++(0.75,-0.55) node[right] {$p_6$};
  \foreach \p in {T,B,L1,L2,L3,R1,R2,R3} \node[v] at (\p) {};
\end{tikzpicture}
\caption{Schematic double-pentagon topology used for the physical-region DP
transport benchmark.}
\label{fig:dp-physical-topology}
\end{figure}
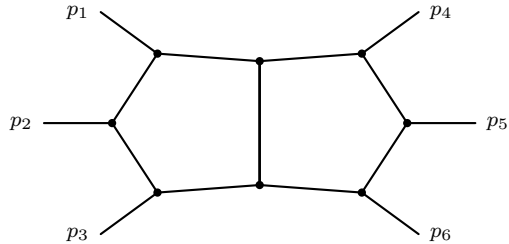

Table~\ref{tab:dp-physical} shows that the released sequential-weight
\textsc{Chess} implementation reproduces the high-precision transport behavior
observed in Ref.~\cite{Liu:2026TwoLoopSixPoint}.  With only
24 nodes the maximum discrepancy with respect to the \textsc{AMFlow} reference
is already \(1.37\times 10^{-31}\).  The error decreases steadily through
72 nodes, reaching \(3.47\times10^{-93}\), and then saturates at the
\(10^{-101}\) level for the 96- and 120-node runs.  This last plateau is
consistent with the finite precision of the independent \textsc{AMFlow}
reference values rather than with a loss of spectral convergence.

\begin{table}[H]
\centering
\caption{Physical-region DP transport from \(\mathbf{x}_0\) to \(\mathbf{x}_1\),
checked against the independent \textsc{AMFlow} value at \(\mathbf{x}_1\).  All
reported runs use 24 requested and actual parallel kernels,
\texttt{"Precision"} \(=160\), \texttt{"WorkingPrecisionA"} \(=220\), and no
endpoint regularization.}
\label{tab:dp-physical}
\footnotesize
\vspace{5pt}
\begin{tabular*}{\linewidth}{@{\extracolsep{\fill}}rrrr@{}}
\toprule
Nodes & Time (s) & RSS (MB) & $\|J-J_{\rm AMFlow}\|_\infty$ \\
\midrule
24 & 8.78 & 4600.3 & $1.37\times 10^{-31}$ \\
48 & 11.82 & 4753.9 & $1.96\times 10^{-62}$ \\
72 & 14.74 & 4827.1 & $3.47\times 10^{-93}$ \\
96 & 18.25 & 4953.5 & $1.12\times 10^{-101}$ \\
120 & 22.54 & 5027.9 & $1.12\times 10^{-101}$ \\
\bottomrule
\end{tabular*}

\end{table}

\subsection{Endpoint-regularized three-loop five-point benchmarks}
\label{subsec:3l5p-regularized}

The second benchmark uses the four planar massless three-loop five-point
families PBB, BPB, BHB, and PBP from Ref.~\cite{Liu:2024ont,
Chicherin:2025ThreeLoopFivePointComplete}.  The families are represented by
their canonical differential-equation matrices and numerical boundary values
at the maximally symmetric Euclidean point.  The transport line goes to a
nearby Euclidean point.  In this setup the symmetric point produces a spurious
regular singularity in the pulled-back matrix, so the left endpoint is
regularized by the lifted spectral formulation described in
Sec.~\ref{sec:method}.

\begin{figure}[H]
\centering
\begin{tabular}{cc}
\subfigure[PBB]{
\begin{tikzpicture}[thick, scale=0.54]
  \coordinate (1) at (4.59314,0);
  \coordinate (2) at (5.75464,1.52085);
  \coordinate (3) at (4.59678,3.04424);
  \coordinate (4) at (0.00141206,2.44223);
  \coordinate (5) at (0,0.602005);
  \coordinate (6) at (4.14059,0.785867);
  \coordinate (7) at (4.83428,1.52169);
  \coordinate (8) at (4.14225,2.25834);
  \coordinate (9) at (0.771176,1.95129);
  \coordinate (10) at (0.770847,1.0941);
  \coordinate (11) at (3.00879,1.09988);
  \coordinate (12) at (1.85009,1.89538);
  \coordinate (13) at (3.00904,1.94518);
  \coordinate (14) at (1.85063,1.15019);
  \draw (8)--(13)(9)--(12)(11)--(13)(12)--(14)(6)--(11)(9)--(10)(11)--(14)(7)--(8)(12)--(13)(10)--(14)(6)--(7);
  \draw (1)--(6) (2)--(7)(3)--(8)(4)--(9)(5)--(10);
  \node[above] at (5.99,1.22) {\small $2$};
  \node[above] at (4.8871,2.85158) {\small $3$};
  \node[above] at (0.386282,2.3) {\small $4$};
  \node[right] at (4.6,0) {\small $1$};
  \node[below] at (0.385335,0.7) {\small $5$};
\end{tikzpicture}
\label{fig:PBB}}
&
\subfigure[BPB]{
\begin{tikzpicture}[thick, scale=0.57]
  \coordinate (1) at (0.542344,0);
  \coordinate (2) at (0,1.76208);
  \coordinate (3) at (2.86764,3.22052);
  \coordinate (4) at (5.73463,1.76188);
  \coordinate (5) at (5.19173,0.000440965);
  \coordinate (6) at (1.17398,0.661902);
  \coordinate (7) at (0.874812,1.4885);
  \coordinate (8) at (2.86652,2.31261);
  \coordinate (9) at (4.85855,1.48833);
  \coordinate (10) at (4.55936,0.661564);
  \coordinate (11) at (2.2473,0.935929);
  \coordinate (12) at (3.48697,0.935849);
  \coordinate (13) at (1.90547,1.7419);
  \coordinate (14) at (3.82728,1.74175);
  \draw (8)--(13)(7)--(13)(9)--(14)(11)--(13)(12)--(14)(6)--(11)(9)--(10)(11)--(12)(8)--(14)(6)--(7)(10)--(12);
  \draw (1)--(6) (2)--(7)(3)--(8)(4)--(9)(5)--(10);
  \node[above] at (5.3,1.7) {\small $2$};
  \node[above] at (3.2,2.9) {\small $3$};
  \node[above] at (0.3,1.7) {\small $4$};
  \node[below] at (5.4,0.7) {\small $1$};
  \node[below] at (0.3,0.8) {\small $5$};
\end{tikzpicture}
\label{fig:BPB}}
\\[1.4em]
\subfigure[BHB]{
\begin{tikzpicture}[thick, scale=0.60]
  \coordinate (1) at (0,3.3263);
  \coordinate (2) at (3.39874,3.37806);
  \coordinate (3) at (4.43267,1.68833);
  \coordinate (4) at (3.39699,0);
  \coordinate (5) at (0.000346146,0.0515417);
  \coordinate (6) at (0.658947,2.67226);
  \coordinate (7) at (2.90173,2.60935);
  \coordinate (8) at (3.51164,1.68889);
  \coordinate (9) at (2.90067,0.7698);
  \coordinate (10) at (0.658598,0.706904);
  \coordinate (11) at (1.36335,1.68989);
  \coordinate (12) at (0.592501,1.69016);
  \coordinate (13) at (1.71772,2.53747);
  \coordinate (14) at (1.71696,0.842312);
  \draw (7)--(13)(9)--(14)(11)--(13)(6)--(13)(11)--(12)(11)--(14)(7)--(8)(10)--(14)(8)--(9)(10)--(12)(6)--(12);
  \draw (1)--(6) (2)--(7)(3)--(8)(4)--(9)(5)--(10);
  \node[] at (4.2,2) {\small $3$};
  \node[] at (3.6,3.2) {\small $4$};
  \node[] at (0.7,3.2) {\small $5$};
  \node[] at (3.6,0.4) {\small $2$};
  \node[] at (0.7,0.2) {\small $1$};
\end{tikzpicture}
\label{fig:BHB}}
&
\subfigure[PBP]{
\begin{tikzpicture}[thick, scale=0.60]
  \coordinate (1) at (0.8166,0.000709659);
  \coordinate (2) at (0,1.78333);
  \coordinate (3) at (2.51243,3.9827);
  \coordinate (4) at (5.02661,1.78427);
  \coordinate (5) at (4.20993,0);
  \coordinate (6) at (1.37367,0.733304);
  \coordinate (7) at (0.914117,1.6497);
  \coordinate (8) at (2.51307,3.0549);
  \coordinate (9) at (4.11155,1.65009);
  \coordinate (10) at (3.65257,0.734098);
  \coordinate (11) at (2.51315,1.83164);
  \coordinate (12) at (1.78218,2.32175);
  \coordinate (13) at (3.24285,2.32174);
  \coordinate (14) at (2.51275,0.889231);
  \draw (8)--(13)(6)--(14)(9)--(13)(11)--(13)(7)--(12)(9)--(10)(11)--(12)(8)--(12)(11)--(14)(10)--(14)(6)--(7);
  \draw (1)--(6) (2)--(7)(3)--(8)(4)--(9)(5)--(10);
  \node[above] at (4.7,1.7) {\small $2$};
  \node[above] at (2.8,3.5) {\small $3$};
  \node[above] at (0.3,1.7) {\small $4$};
  \node[below] at (4.4,0.8) {\small $1$};
  \node[below] at (0.7,0.8) {\small $5$};
\end{tikzpicture}
\label{fig:PBP}}
\end{tabular}
\caption{Genuine three-loop five-point integral families: Pentagon--Box--Box
(PBB), Box--Pentagon--Box (BPB), Box--Hexagon--Box (BHB), and
Pentagon--Box--Pentagon (PBP).  The diagrams follow the topology convention of
Ref.~\cite{Chicherin:2025ThreeLoopFivePointComplete}.}
\label{fig:3l5p-topologies}
\end{figure}
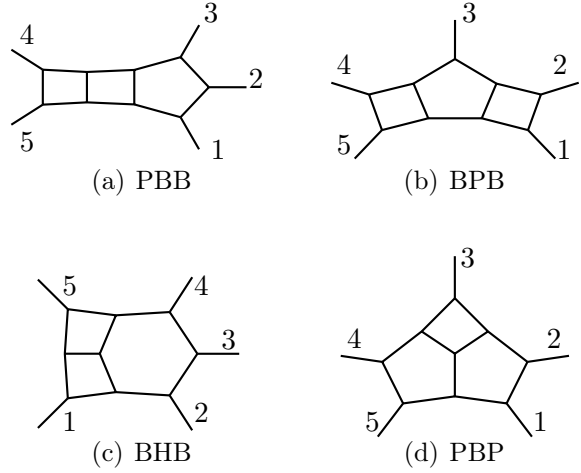

Let \(F_m\) denote the complete all-weight final-state matrix obtained with
\(m\) Chebyshev subintervals.  Table~\ref{tab:3l5p} reports the
node-refinement estimate
\(\Delta_{\rm node}=\|F_{120}-F_{96}\|_\infty\).  In the absence of external
high-precision reference values for all four families in this benchmark setup,
this quantity is used as
an internal convergence diagnostic.  It compares two separately assembled
Chebyshev node refinements of the same transport problem and should be
interpreted as an absolute max-norm stability estimate, not as a relative error
bound for each individual integral.  The node-difference norms correspond to
about 70 absolute digits.  The largest PBP system has 734 master integrals and
still completes in less than 80 seconds at 120 nodes with a process-tree peak
RSS below 9 GB.

\begin{table}[H]
\centering
\caption{Three-loop five-point benchmark with left-endpoint regularization.
RSS denotes process-tree peak resident memory.  All rows use Mathematica 14.3,
24 requested and actual parallel kernels, \texttt{"Precision"} \(=100\), and
\texttt{"WorkingPrecisionA"} \(=120\).}
\label{tab:3l5p}
\footnotesize
\vspace{5pt}
\begin{tabular*}{\linewidth}{@{\extracolsep{\fill}}lrrrrr@{}}
\toprule
Family & $t_{96}$ (s) & $t_{120}$ (s) & RSS$_{120}$ (MB) & $\|J_{120}-J_{96}\|_\infty$ & digits \\
\midrule
PBB & 15.05 & 18.83 & 5062.6 & $3.56\times 10^{-71}$ & 70.4 \\
BPB & 20.63 & 23.16 & 5368.2 & $3.76\times 10^{-71}$ & 70.4 \\
BHB & 23.41 & 27.94 & 5487.5 & $4.89\times 10^{-71}$ & 70.3 \\
PBP & 65.64 & 76.49 & 8656.0 & $5.64\times 10^{-71}$ & 70.2 \\
\bottomrule
\end{tabular*}

\end{table}

We also include a fixed-protocol comparison with \textsc{DiffExp}~\cite{Hidding:2020ytt}
on the same four three-loop five-point families.  This comparison is intended
as a reproducible workflow benchmark on the prepared straight-line transports,
not as a universal performance statement about \textsc{DiffExp} or about
local-series methods.  Different path subdivisions, expansion orders, Padé
settings, or problem-specific tuning may change the performance of a
local-series workflow.  The canonical matrices were converted from the same
auxiliary files into the \textsc{DiffExp} one-line matrix format, and the
transport used the same straight path, the same finite boundary values, and 24
requested parallel kernels.  The \textsc{DiffExp} settings were
\texttt{WorkingPrecision} \(=80\), \texttt{ExpansionOrder} \(=20\),
\texttt{AccuracyGoal} \(=15\), \texttt{ChopPrecision} \(=40\), and Padé
evaluation enabled.  We report the observed final-state agreement rather than
interpreting the internal \texttt{AccuracyGoal} parameter as a direct prediction
of final digits.

Table~\ref{tab:3l5p-diffexp} compares the 120-node \textsc{Chess} runs with
these \textsc{DiffExp} runs.  The two accuracy columns are diagnostics of
different type.  The \textsc{Chess} value is
\(d_{\rm node}=-\log_{10}\|F^{\rm Chess}_{120}-F^{\rm Chess}_{96}\|_\infty\), while the
\textsc{DiffExp} value is
\(d_{\rm cross}=-\log_{10}\|F_{\rm DiffExp}-F^{\rm Chess}_{120}\|_\infty\).
The latter is a cross-method consistency check against the current
\textsc{Chess} final state, not an independent proof of the absolute accuracy
of either method.  With this fixed protocol, \textsc{Chess} is faster by
factors between \(8.8\) and \(17.9\).  The process-tree RSS includes the full
24-kernel Wolfram process tree, so the fixed subkernel baseline is visible for
the smaller families.  For the largest PBP system,
where the algorithmic memory footprint is visible above this baseline,
\textsc{Chess} uses less memory in this run.

\begin{table}[H]
\centering
\caption{\textsc{Chess} and \textsc{DiffExp} comparison for the four
three-loop five-point families.  RSS is the process-tree peak resident memory.
The accuracy columns are diagnostics of different type:
\(d_{\rm node}\) is the \textsc{Chess} node-refinement estimate, while
\(d_{\rm cross}\) is the \textsc{DiffExp} final-state difference with respect
to the current \textsc{Chess} \(n=120\) result.}
\label{tab:3l5p-diffexp}
\footnotesize
\vspace{5pt}
\begin{tabular*}{\linewidth}{@{\extracolsep{\fill}}lrrrrrrr@{}}
\toprule
Family & masters & \multicolumn{2}{c}{time (s)} & \multicolumn{2}{c}{RSS (MB)} & \multicolumn{2}{c}{digits} \\
\cmidrule(lr){3-4}\cmidrule(lr){5-6}\cmidrule(lr){7-8}
 & & \textsc{Chess} & \textsc{DiffExp} & \textsc{Chess} & \textsc{DiffExp} & \textsc{Chess} & \textsc{DiffExp} \\
\midrule
PBB & 316 & 18.83 & 166.36 & 5062.6 & 4784.3 & 70.4 & 45.9 \\
BPB & 367 & 23.16 & 226.24 & 5368.2 & 4910.9 & 70.4 & 46.6 \\
BHB & 431 & 27.94 & 316.72 & 5487.5 & 5047.8 & 70.3 & 46.4 \\
PBP & 734 & 76.49 & 1365.50 & 8656.0 & 13026.6 & 70.2 & 58.4 \\
\bottomrule
\end{tabular*}

\end{table}

\begin{figure}[H]
\centering
\includegraphics[width=\linewidth]{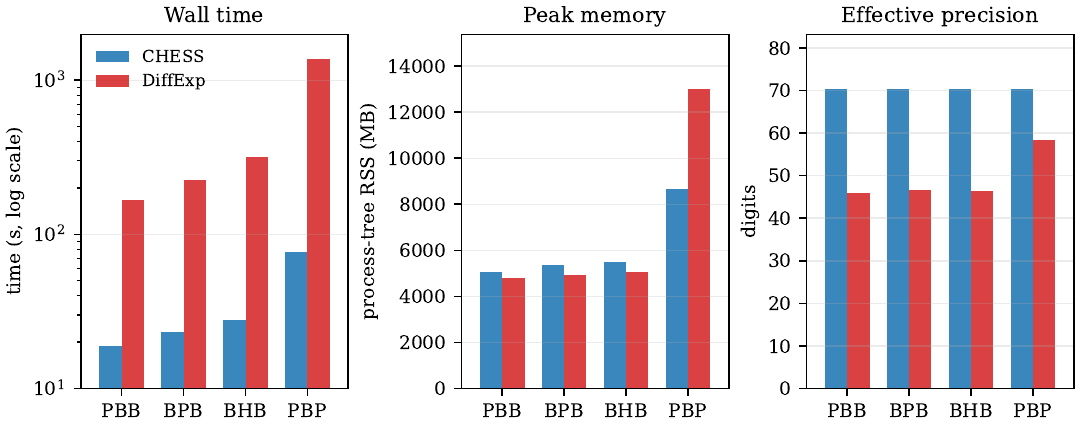}
\caption{Grouped-bar comparison of wall time, process-tree peak RSS, and
accuracy diagnostics for the four three-loop five-point families.  For
\textsc{Chess} the accuracy diagnostic is the 120-versus-96 node difference;
for \textsc{DiffExp} it is the final-state difference with respect to the
current \textsc{Chess} 120-node result.  The time panel uses a logarithmic
vertical scale.}
\label{fig:3l5p-diffexp-comparison}
\end{figure}

\subsection{Direct-matrix benchmark for an elliptic case}
\label{subsec:np-triangle-direct}

The last benchmark uses family (b) of the two-loop non-planar triangle systems
studied in Ref.~\cite{Jiang:2023jmk}.  The published canonical differential
equation is \(\eps\)-factorized, while its one-dimensional coefficient matrix
contains elliptic periods.  The test checks the direct evaluator interface in
a case where the
entries of \(B(y)\) are not just rational functions or derivatives of
ordinary logarithms.

The family depends on
\begin{equation}
  y=-\frac{m^2}{s},
  \label{eq:nptriangle-y-variable}
\end{equation}
with two massive internal lines and two massless external legs.  The elliptic
sector is controlled by a quartic curve.  Ref.~\cite{Jiang:2023jmk} chooses a
period \(\psi_0(y)\), a Wronskian \(W(y)\), and the modular parameter
\(q=\exp(2\pi i\tau)\).  In our direct evaluator the only elliptic special
function evaluated at the Chebyshev nodes is the complete elliptic integral of
the first kind,
\begin{equation}
  \psi_0(y)=
  \frac{2}{\pi}\,
  \frac{y^2 K(k^2)}
       {\sqrt{(u_3-u_1)(u_4-u_2)}} ,
  \qquad
  k^2 =
  \frac{1-4y-8y^2-\sqrt{1-8y}}
       {1-4y-8y^2+\sqrt{1-8y}},
  \label{eq:nptriangle-elliptic-period}
\end{equation}
where \(K(m)\) denotes the complete elliptic integral of the first kind with
parameter \(m\), the same convention used by \texttt{EllipticK[m]} in
\textsc{Mathematica}.  The branch points are
\(u_1=-1\), \(u_2=-(\sqrt{1-8y}+1)^2/4\),
\(u_3=-(\sqrt{1-8y}-1)^2/4\), and \(u_4=0\).  No complete elliptic integral of
the second kind is used in this example.
In the normalization used by the auxiliary files, the canonical differential
equation for family (b) can be written as
\begin{equation}
  J(y)\frac{\dd}{\dd y}\vecJ(y,\eps)=\eps\,A(y)\vecJ(y,\eps),
  \qquad
  J(y)=\frac{\psi_0(y)^2}{W(y)} .
  \label{eq:nptriangle-author-system}
\end{equation}
We pass the direct matrix
\begin{equation}
  B(y)=\frac{A(y)}{J(y)}
  \label{eq:nptriangle-direct-matrix}
\end{equation}
to \textsc{Chess}.  The entries include rational functions, square roots, and
the period \(\psi_0(y)\) in Eq.~\eqref{eq:nptriangle-elliptic-period}.  No
conversion to a dlog letter list is made.  The code evaluates \(B(y)\) at the
Chebyshev nodes with the requested matrix precision and then applies the same
spectral propagator used in the previous benchmarks.

\begin{figure}[H]
\centering
\includegraphics[width=0.48\linewidth]{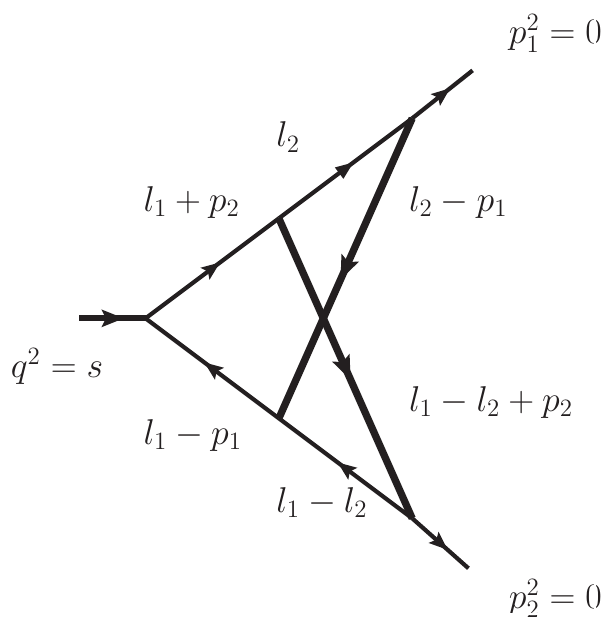}
\caption{Two-loop non-planar triangle family (b) of
Ref.~\cite{Jiang:2023jmk}.  Thick internal lines denote propagators with mass
\(m\); the other internal lines are massless.}
\label{fig:np-triangle-family-b}
\end{figure}

For the numerical test we use the closed 15-dimensional sub-sector of the
published 18-dimensional canonical system.  We transport the five coefficients
from \(\eps^0\) through \(\eps^4\) along
\begin{equation}
  y:\frac{1}{100}\longrightarrow \frac{1}{20},
  \qquad
  s:-100\longrightarrow -20,
  \label{eq:nptriangle-path}
\end{equation}
without endpoint regularization.  Both endpoints lie in the Euclidean region
of this one-scale problem.  The boundary and reference values were computed
independently with \textsc{AMFlow}, using Kira reduction, 24 threads,
\texttt{precision=200}, and \texttt{epsOrder=10}.  The two endpoint runs took
\(710.11\) s and \(564.30\) s.

Table~\ref{tab:np-triangle-direct} gives the node-refinement behavior against
the 200-digit \textsc{AMFlow} endpoint at \(y=1/20\).  The error decreases
geometrically until it reaches the available reference precision.  At 448
nodes the maximum absolute difference is about \(10^{-186}\).  At 512 nodes
the printed difference is below \(10^{-200}\), which should be read as
saturation against the 200-digit reference rather than as an independently
certified 208-digit result.

\begin{table}[H]
\centering
\caption{Two-loop non-planar triangle family (b) from \(y=1/100\) to
\(y=1/20\), checked against an independent 200-digit \textsc{AMFlow} endpoint.
All rows use 24 requested and actual parallel kernels, no endpoint
regularization, and \(\eps^0,\ldots,\eps^4\).  The error is the maximum
absolute difference over the 15-dimensional closed sub-sector.}
\label{tab:np-triangle-direct}
\footnotesize
\vspace{5pt}
\begin{tabular*}{\linewidth}{@{\extracolsep{\fill}}rrrrr@{}}
\toprule
Nodes & Precision & WorkingPrecisionA & Time (s) & $\|J-J_{\rm ref}\|_\infty$ \\
\midrule
96  & 260 & 360 & 3.61  & $3.50\times 10^{-39}$ \\
144 & 260 & 360 & 0.89  & $2.40\times 10^{-59}$ \\
192 & 292 & 392 & 1.83  & $1.75\times 10^{-79}$ \\
240 & 340 & 440 & 3.11  & $1.32\times 10^{-99}$ \\
320 & 420 & 520 & 8.17  & $4.02\times 10^{-133}$ \\
384 & 484 & 584 & 19.84 & $6.36\times 10^{-160}$ \\
448 & 548 & 648 & 26.62 & $1.02\times 10^{-186}$ \\
512 & 612 & 712 & 44.56 & $1.01\times 10^{-208}$ \\
\bottomrule
\end{tabular*}

\end{table}

\section{Conclusions}
\label{sec:conclusions}

We have presented \textsc{Chess}, a Wolfram Language package for Chebyshev
pseudo-spectral transport of \(\eps\)-factorized one-dimensional differential
equations for Feynman master integrals.
The prepared-data workflow assembles the path dependence from scalar pullbacks
of differential one-forms and sparse constant matrices.  The core propagator
only requires a numerical evaluator for the pulled-back matrix \(B(t)\), and
can therefore also be used with direct matrix evaluators, as illustrated by the
non-planar triangle benchmark.

The method uses structure that is already present in the transport problem.
After a path is chosen, the system is one-dimensional.  The
\(\eps\)-factorized form turns the retained Laurent coefficients into a
triangular hierarchy, so each coefficient is obtained from the previous one
with the same collocation operator.  In the prepared-data workflow, the path
dependence is separated from the constant slot matrices, allowing sparse matrix
data to be cached and reused.  For spurious regular singular endpoints,
\textsc{Chess} replaces the singular endpoint equation by a finite
residue/finite-part condition, either through a right-endpoint row replacement
or through a lifted left-endpoint unknown.

The benchmarks show the behavior expected from a spectral method on analytic
path segments.  The two-loop six-point double-pentagon example agrees with
independent \textsc{AMFlow} reference data to the available reference precision.
The four three-loop five-point examples test large endpoint-regularized systems
and show stable node-refinement differences at about 70 absolute digits in the
reported max norm.  A fixed-protocol comparison with \textsc{DiffExp} on the
same straight-line transports provides an additional cross-method check and
gives shorter wall times for the options used here, with
comparable memory use in the smaller parallel runs and lower process-tree RSS
for the largest system.  The non-planar triangle example shows that the core
collocation engine is not restricted to the prepared-data assembly layer once
a suitable numerical matrix evaluator is supplied.  In this test, the
independent \textsc{AMFlow} endpoint calculations take about 21 minutes,
whereas the \textsc{Chess} transports take seconds and reach the effective
precision limit of the 200-digit reference values.

The present release is intentionally focused.  It assumes that the basis,
boundary constants, path, and analytic branch have already been specified by
the user.  It does not perform integral reduction, construct canonical or
\(\eps\)-factorized bases, determine boundary constants, or automatically
choose paths around singularities.  These restrictions are important: Chebyshev
convergence depends on analyticity of the supplied coefficient functions in a
complex neighborhood of the path, and endpoint regularization applies only to
spurious regular singular behavior for the selected finite branch.

Several extensions would be useful in practice: automated path diagnostics,
estimates of the nearest complex singularities controlling the Bernstein
convergence parameter, more systematic branch tracking, direct user-supplied
residue data for direct-matrix endpoint regularization, and interfaces to
external reduction and boundary-value tools.  The current package supplies the
transport layer after those analytic preparations have been completed.

\section*{Code and data availability}

The accompanying compact program archive is intended to contain the
\textsc{Chess} package file, runnable examples, the data files needed by those
examples, sample input/output files, and a short README.  Large development
artifacts and full benchmark-generation scripts are not part of the compact
user-facing release.  The CPC Program Library link will be assigned by the
Technical Editor after acceptance.  The developer repository is
\url{https://github.com/Alice-Shimada/CHESS}.  The benchmark tables and figure
in this manuscript are generated from machine-readable files in the manuscript
benchmark workspace.  The fixed-protocol \textsc{DiffExp} comparison and the
non-planar triangle direct-matrix benchmark are documented as manuscript-side
reproducibility material rather than as part of the minimal user archive.

\section*{Acknowledgments}
We warmly thank Dmitry Chicherin,  Yu Wu, Yingxuan Xu and Yongqun Xu for fruitful discussion. We also thank Johannes Henn for hospitality in Max Planck Institute for Physics where part of this work has been accomplished. The research of Yuanche Liu is supported by NSFC through Grant No. 124B1014. Yang Zhang is supported by NSFC through Grant No. 12575078 and No. 12247103. 
\appendix
\section{Endpoint regularization details}
\label{app:endpoint}

This appendix gives the algebra behind the endpoint formulas used in
Sec.~\ref{subsec:endpoint}.  The discussion is local near one endpoint and
uses the global transport coordinate \(t\).  Let \(c\) be either endpoint and
assume that the pulled-back matrix has a regular singular expansion
\begin{equation}
  B(t)=\frac{R_c}{t-c}+M_c+O(t-c),
  \label{eq:app-B-laurent}
\end{equation}
with constant matrices \(R_c\) and \(M_c\).  The expansion is one-sided on the
transport interval.  At the right endpoint the implementation may compute the
series in \(s=b-t\), but the stored residue is converted to the convention of
Eq.~\eqref{eq:app-B-laurent}.

\subsection{Finite endpoint source}
\label{appsubsec:finite-source}

For the \(\eps\)-coefficient hierarchy,
\begin{equation}
  \frac{\dd}{\dd t}\vecJ^{(r)}(t)
    =
  B(t)\vecJ^{(r-1)}(t),
  \qquad r\ge 1,
  \label{eq:app-hierarchy}
\end{equation}
we use the non-negative order index obtained after shifting the lowest retained
Laurent coefficient to \(r=0\).  The lowest coefficient has zero source.
The only singular term in the right-hand side near \(c\) is
\begin{equation}
  \frac{R_c\vecJ^{(r-1)}(c)}{t-c}.
  \label{eq:app-singular-source-term}
\end{equation}
For a spurious endpoint singularity, the finite solution branch satisfies
\begin{equation}
  R_c\vecJ^{(r)}(c)=0
  \label{eq:app-regularity}
\end{equation}
at all orders represented in the boundary data.  Applying this condition to
\(\vecJ^{(r-1)}\) removes the pole in
Eq.~\eqref{eq:app-hierarchy}.  The remaining finite value is obtained by
expanding
\begin{equation}
  \vecJ^{(r-1)}(t)
    =
  \vecJ^{(r-1)}(c)
    +(t-c)\vecJ^{(r-1)\prime}(c)
    +O((t-c)^2).
  \label{eq:app-J-expand}
\end{equation}
Substitution gives
\begin{equation}
  \lim_{t\to c}
    B(t)\vecJ^{(r-1)}(t)
  =
  M_c\vecJ^{(r-1)}(c)
    +R_c\vecJ^{(r-1)\prime}(c).
  \label{eq:app-finite-limit}
\end{equation}
Writing
\begin{equation}
  S_r(c)=\vecJ^{(r)\prime}(c)
  \label{eq:app-source-def}
\end{equation}
for the finite endpoint derivative source, Eq.~\eqref{eq:app-finite-limit}
becomes the recursion
\begin{equation}
  S_0(c)=0,\qquad
  S_r(c)
    =
  M_c\vecJ^{(r-1)}(c)+R_cS_{r-1}(c),
  \qquad r\ge 1.
  \label{eq:app-source-recursion}
\end{equation}
Iterating gives the equivalent explicit form
\begin{equation}
  S_r(c)
    =
  \sum_{q=1}^{r}
    R_c^{q-1}M_c\vecJ^{(r-q)}(c).
  \label{eq:app-source-explicit}
\end{equation}
The recursive form is the one used in the implementation because it only
requires one sparse multiplication by \(M_c\) and one sparse multiplication by
\(R_c\) per order.

The order of multiplication matters for efficiency.  In many canonical
systems \(M_c\vecJ^{(r-q)}(c)\) is much sparser than a generic vector,
because the boundary vector and the finite endpoint operator satisfy many
structural relations.  For this reason \textsc{Chess} first forms the vector
\(M_c\vecJ^{(r-q)}(c)\) and only then applies the residue powers.  It
does not precompute dense or less-sparse composite matrices such as
\(R_c^{q-1}M_c\).  This is also the reason that the endpoint operator is
assembled from the same scalar-pullback slots used in \(B(t)\), instead of
first collapsing all slots with equal endpoint weights into a large
intermediate matrix.

\subsection{Right endpoint row replacement}
\label{appsubsec:right-row}

Let \(t_0,\ldots,t_m\) be the increasing Chebyshev--Lobatto grid and let
\(D\) be the corresponding differentiation matrix.  If only the right endpoint
\(b=t_m\) is regularized, the unknowns remain the values
\(\vecJ^{(r)}(t_j)\).  The ordinary collocation equations are used at
the interior nodes,
\begin{equation}
  \sum_{\ell=0}^{m}D_{j\ell}\vecJ^{(r)}(t_\ell)
    =
  B(t_j)\vecJ^{(r-1)}(t_j),
  \qquad j=1,\ldots,m-1,
  \label{eq:app-right-interior}
\end{equation}
while the endpoint equation is replaced by
\begin{equation}
  \sum_{\ell=0}^{m}D_{m\ell}\vecJ^{(r)}(t_\ell)
    =
  S_r(b).
  \label{eq:app-right-row}
\end{equation}
The boundary row at \(t_0=a\) remains
\(\vecJ^{(r)}(a)=\vecJ^{(r)}_a\).  Thus the scalar node operator is
the ordinary Chebyshev differentiation operator with the first row replaced by
the left boundary condition and the last row interpreted as the finite
endpoint derivative.  No value of \(B(b)\) is ever needed.

\subsection{Left endpoint lifting}
\label{appsubsec:left-lift}

At the left endpoint \(a=t_0\), the value
\(\vecJ^{(r)}(a)=\vecJ^{(r)}_a\) is part of the input data.  If the
unknowns were kept as the values \(\vecJ^{(r)}(t_j)\), the first row of
the collocation system would already be used to impose this boundary value.
There would be no separate row at the same node on which to place the finite
endpoint derivative.  The left-regularized system is therefore written in terms
of the difference-quotient unknown
\begin{equation}
  \vecJ^{(r)}(t)
    =
  \vecJ^{(r)}_a+(t-a)\mathbf{Z}^{(r)}(t),
  \label{eq:app-left-lift}
\end{equation}
or, for \(t>a\),
\begin{equation}
  \mathbf{Z}^{(r)}(t)
    =
  \frac{\vecJ^{(r)}(t)-\vecJ^{(r)}_a}{t-a}.
  \label{eq:app-left-quotient}
\end{equation}
For the regular branch, this quotient has a finite one-sided limit at
\(a\), and that limit is the endpoint derivative
\(\vecJ^{(r)\prime}(a)\).  Differentiating
Eq.~\eqref{eq:app-left-lift} gives
\begin{equation}
  \vecJ^{(r)\prime}(t)
    =
  \mathbf{Z}^{(r)}(t)+(t-a)\mathbf{Z}^{(r)\prime}(t).
  \label{eq:app-left-derivative}
\end{equation}
For each node \(t_j\), define the scalar lifted collocation operator
\begin{equation}
  L_{j\ell}
    =
  \delta_{j\ell}
    +(t_j-a)D_{j\ell}.
  \label{eq:app-lifted-operator}
\end{equation}
This operator is the Chebyshev representation of
\(\mathbf{Z}^{(r)}+(t-a)\mathbf{Z}^{(r)\prime}\).  The differential equation is
then imposed at all non-left endpoint rows as
\begin{equation}
  \sum_{\ell=0}^{m}L_{j\ell}\mathbf{Z}^{(r)}(t_\ell)
    =
  B(t_j)\vecJ^{(r-1)}(t_j),
  \qquad j=1,\ldots,m,
  \label{eq:app-left-interior}
\end{equation}
when no right endpoint regularization is requested.  The left endpoint row is
not obtained by substituting \(t=a\) into \(B(t)\).  Instead, it fixes the
finite one-sided derivative,
\begin{equation}
  \mathbf{Z}^{(r)}(a)=S_r(a).
  \label{eq:app-left-row}
\end{equation}
Here \(S_r(a)\) is computed by the residue recursion
Eq.~\eqref{eq:app-source-recursion}, using the already reconstructed lower
order values at \(a\).  If the right endpoint is also regularized,
Eq.~\eqref{eq:app-left-interior} is imposed only for \(j=1,\ldots,m-1\), and
the final row is replaced by
\begin{equation}
  \sum_{\ell=0}^{m}L_{m\ell}\mathbf{Z}^{(r)}(t_\ell)
    =
  S_r(b).
  \label{eq:app-both-endpoints-row}
\end{equation}
Here \(S_r(b)\) is computed from the already reconstructed lower-order values
\(\vecJ^{(q)}(b)\), \(q<r\), using the right-endpoint residue and finite
matrices.
After the lifted system has been solved, the original coefficient values are
recovered by
\begin{equation}
  \vecJ^{(r)}(t_j)
    =
  \vecJ^{(r)}_a+(t_j-a)\mathbf{Z}^{(r)}(t_j).
  \label{eq:app-left-recover}
\end{equation}
This transformation keeps the solve in the Chebyshev collocation framework.
The unknown has changed from endpoint values of \(\vecJ^{(r)}\) to endpoint
values of \(\mathbf{Z}^{(r)}\), and the scalar node operator has changed from
\(D\) to \(L\).  The matrix data \(B(t_j)\), \(R_a\), and \(M_a\) still enter
only through the ordinary right-hand side and the finite endpoint source.

For \(r=0\), the hierarchy gives a constant solution.  In the lifted left
endpoint formulation this corresponds to a zero right-hand side and
\(\mathbf{Z}^{(0)}=0\), consistent with \(S_0(a)=0\).  Higher orders are then
computed sequentially, because the right-hand side for order \(r\) uses the
already reconstructed values of order \(r-1\).

\subsection{Extraction of residue and finite matrices from scalar pullbacks}
\label{appsubsec:residue-extraction}

The endpoint matrices are assembled from the same prepared decomposition as
the ordinary matrix \(B(t)\).  Suppose
\begin{equation}
  B(t)=\sum_{\alpha}A_{\alpha} b_{\alpha}(t).
  \label{eq:app-prepared-pullback-decomposition}
\end{equation}
Near an endpoint,
\begin{equation}
  b_{\alpha}(t)
    =
  \frac{r_{\alpha,c}}{t-c}
    +m_{\alpha,c}
    +O(t-c).
  \label{eq:app-pullback-laurent}
\end{equation}
Then
\begin{equation}
  R_c=\sum_{\alpha}A_{\alpha} r_{\alpha,c},
  \qquad
  M_c=\sum_{\alpha}A_{\alpha} m_{\alpha,c}.
  \label{eq:app-RM-pullback}
\end{equation}
Scalar pullbacks that are finite at the endpoint have \(r_{\alpha,c}=0\), and
their finite coefficient is obtained by direct endpoint substitution.  Only
pullbacks that fail direct substitution require Laurent coefficients, which
are computed with \texttt{SeriesCoefficient}.  This avoids repeated residue
extraction from a full matrix expression and preserves the sparse slot
structure of the decomposed system.

The construction assumes that the one-sided limits in
Eq.~\eqref{eq:app-pullback-laurent} exist at the requested working precision and
that the supplied boundary data satisfy Eq.~\eqref{eq:app-regularity}.  If
these assumptions fail, endpoint regularization is not appropriate for that
transport path or boundary point.

\section{Reproducibility details}
\label{app:reproducibility}

The manuscript workspace separates package code, runnable examples, curated
results, and paper-facing tables.  This appendix documents that manuscript
workspace, not the compact user-facing program archive described in
App.~\ref{app:program-archive}.  The package loaded by all reported Chebyshev
runs is
\begin{verbatim}
package/Chess.wl
\end{verbatim}
The example and benchmark drivers live under \path{scripts/}, and the
machine-readable outputs used to generate the tables live under
\path{results/}.  The LaTeX tables in \path{paper/tables/} are generated
from these curated result files rather than edited by hand.

\subsection{Chebyshev benchmark generation}

The three-loop five-point benchmarks in Table~\ref{tab:3l5p} were regenerated
with
\begin{verbatim}
scripts/run_latest_chess_benchmarks.sh
wolframscript -file scripts/collect_latest_results.wls
\end{verbatim}
The shell runner calls the Mathematica drivers
\begin{verbatim}
scripts/run_chess_3l5p_family.wls
\end{verbatim}
with the node counts, precision settings, endpoint options, and 24-kernel
parallel setting used in the paper.  The collector then reads the raw
Mathematica summary files and writes
\begin{verbatim}
results/chess_3l5p_latest.csv
results/chess_3l5p_convergence_latest.csv
paper/tables/chess_3l5p_convergence.tex
\end{verbatim}
The corresponding raw final states and run summaries are stored under
\path{results/raw/}.  These files include the full output matrices used for
node-difference checks.

The \textsc{DiffExp} comparison in Table~\ref{tab:3l5p-diffexp} and
Fig.~\ref{fig:3l5p-diffexp-comparison} uses the separately prepared workspace
\begin{verbatim}
benchmark_supplement/diffexp_comparison/
\end{verbatim}
whose scripts convert the four published three-loop five-point matrices to
\textsc{DiffExp}'s \texttt{d\_1.m} input format and run all four families with
24 requested kernels.  The manuscript-side scripts
\begin{verbatim}
scripts/compare_diffexp_3l5p_against_chess.wls
scripts/make_3l5p_diffexp_comparison.py
\end{verbatim}
then compare the saved \textsc{DiffExp} final states with the current
\textsc{Chess} \(n=120\) final states and generate
\begin{verbatim}
results/diffexp_3l5p_error_vs_chess_n120_wp80.csv
results/diffexp_3l5p_parallel24_wp80.csv
results/chess_diffexp_3l5p_comparison_wp80.csv
paper/tables/chess_diffexp_3l5p_comparison.tex
paper/figures/chess_diffexp_3l5p_comparison.pdf
\end{verbatim}
This comparison workspace is separate from the compact \textsc{Chess} program
archive.  If supplied with the manuscript, it should be treated as benchmark
reproducibility material.  The comparison is a fixed-protocol benchmark of the
stated inputs and options rather than a full exploration of all possible
\textsc{DiffExp} settings.

The physical-region two-loop six-point DP benchmark in Table~\ref{tab:dp-physical}
was generated with
\begin{verbatim}
scripts/run_chess_dp_physical.wls
\end{verbatim}
using the unpacked auxiliary files of Ref.~\cite{Liu:2026TwoLoopSixPoint}.
The exact source directory and auxiliary archive used in the local run are
recorded in the raw Mathematica summary files under \path{results/raw/}.

The non-planar triangle benchmark in Table~\ref{tab:np-triangle-direct} was
generated under
\begin{verbatim}
elliptic_search/larger_candidates/np_triangle_2305_13951/
\end{verbatim}
The endpoint reference values were computed with
\begin{verbatim}
scripts/07_family_b_amflow200_chebyshev_scan.wls
\end{verbatim}
The same script then reuses the two stored 200-digit endpoint files and scans
the Chebyshev node counts.  A separate check against the author's published
canonical \(q\)-expansion is provided by
\begin{verbatim}
scripts/06_family_b_aux_qexp_chess.wls
\end{verbatim}
The direct matrix evaluator is defined in
\path{scripts/triangle_family_b_common.wl}.  The raw \textsc{AMFlow} outputs,
canonical endpoint coefficients, \textsc{Chess} summaries, and logs are stored
under \path{elliptic_search/larger_candidates/np_triangle_2305_13951/}.

\subsection{Memory measurement}

All headline memory numbers in the paper are external process-tree peak RSS
values.  The sampler is
\begin{verbatim}
scripts/monitor_run.sh
\end{verbatim}
It samples the whole Mathematica process tree once per second, including the
main kernel, wrappers, subkernels, and helper processes.  The maximum is
stored in the corresponding memory log.
Mathematica's internal memory counters are kept only as supplementary
diagnostics in the raw summary files.

This convention is important for parallel runs.  A main-kernel memory counter
does not include the resident memory of all subkernels and therefore
understates the memory footprint visible to the operating system.  The
process-tree number is the value used consistently in
Tables~\ref{tab:dp-physical}, \ref{tab:3l5p},
\ref{tab:3l5p-diffexp}, and \ref{tab:np-triangle-direct}.

\subsection{Table checks}

The numerical values printed in the manuscript are taken from the curated CSV
files listed above.  The raw Mathematica output is kept so that every table
entry can be traced back to a particular run log and summary file.

\section{Program archive contents}
\label{app:program-archive}

The public program archive is meant to be smaller than the development
workspace.  It contains the \textsc{Chess} package file, runnable
examples, the data files needed by those examples, sample outputs, a short
README. 
\begin{table}[h]
\centering
\caption{Recommended public archive layout for the \textsc{Chess} program submission.}
\label{tab:archive-layout}
\small
\begin{tabular}{p{0.32\linewidth}p{0.58\linewidth}}
\toprule
Path & Contents \\
\midrule
\path{README.md} &
Short installation and example-running instructions. \\
\path{Chess.wl} &
Main Mathematica package implementing the Chebyshev transport engine. \\
\path{examples/} &
Runnable Mathematica example scripts, pass/fail tests, and sample outputs. \\
\path{data/} &
Prepared matrices, one-dimensional scalar data, boundary values, and reference
values required by the examples. \\
\bottomrule
\end{tabular}
\end{table}

The minimum smoke test is to run
\begin{verbatim}
wolframscript -file examples/run_tests.wls
\end{verbatim}
from the archive root.  The test runs the three public examples used in
the release archive: DP, the four three-loop five-point families
\texttt{PBB}, \texttt{BPB}, \texttt{BHB}, and \texttt{PBP}, and the
non-planar triangle direct-matrix example.  DP and the non-planar triangle
example are checked against bundled independent reference values.  The
three-loop five-point cases are shipped as
endpoint-regularized workflow examples; their release tests check successful
completion and report checksums rather than independent reference errors.  The
sample outputs under \path{examples/expected/} record the expected dimensions,
endpoint diagnostics, and checksums for the release tests.

\bibliographystyle{elsarticle-num}
\bibliography{references}

\end{document}